\documentclass[aps,prd,twocolumn,superscriptaddress]{revtex4-2}
\usepackage[utf8]{inputenc}
\usepackage{graphicx}
\usepackage{xcolor}
\usepackage{longtable}
\usepackage{bm}
\usepackage{amsmath}
\usepackage{comment}
\begin{document}

\title{Direct dark matter searches with the full data set of XMASS-I}
\newcommand{\ICRR}{\affiliation{Kamioka Observatory, Institute for Cosmic Ray Research, the University of Tokyo, Higashi-Mozumi, Kamioka, Hida, Gifu 506-1205, Japan}}
\newcommand{\IBS}{\affiliation{Center for Underground Physics, Institute for Basic Science, 70 Yuseong-daero 1689-gil, Yuseong-gu, Daejeon 305-811, South Korea}}
\newcommand{\ISEE}{\affiliation{Institute for Space-Earth Environmental Research, Nagoya University, Nagoya, Aichi 464-8601, Japan}}
\newcommand{\IPMU}{\affiliation{Kavli Institute for the Physics and Mathematics of the Universe (WPI), the University of Tokyo, Kashiwa, Chiba 277-8582, Japan}}
\newcommand{\SNU}{\affiliation{Department of Physics and Astronomy, Seoul National University, Seoul 151-742, South Korea}}
\newcommand{\KMI}{\affiliation{Kobayashi-Maskawa Institute for the Origin of Particles and the Universe, Nagoya University, Furo-cho, Chikusa-ku, Nagoya, Aichi 464-8602, Japan}}
\newcommand{\Kobe}{\affiliation{Department of Physics, Kobe University, Kobe, Hyogo 657-8501, Japan}}
\newcommand{\KRISS}{\affiliation{Korea Research Institute of Standards and Science, Daejeon 305-340, South Korea}}
\newcommand{\Miyagi}{\affiliation{Department of Physics, Miyagi University of Education, Sendai, Miyagi 980-0845, Japan}}
\newcommand{\Nihon}{\affiliation{Department of Physics, College of Science and Technology, Nihon University, Kanda, Chiyoda-ku, Tokyo 101-8308, Japan}}
\newcommand{\RCNS}{\affiliation{Research Center for Neutrino Scientce, Tohoku Univeristy, Sendai 980-8578, Japan}}
\newcommand{\Tokai}{\affiliation{Department of Physics, Tokai University, Hiratsuka, Kanagawa 259-1292, Japan}}
\newcommand{\Tokushima}{\affiliation{Department of Physics, Tokushima University, 2-1 Minami Josanjimacho Tokushima city, Tokushima 770-8506, Japan}}
\newcommand{\Tsinghua}{\affiliation{Department of Engineering Physics, Tsinghua University, Haidian District, Beijing, China 100084}}
\newcommand{\Waseda}{\affiliation{Waseda Research Institute for Science and Engineering, Waseda University, 3-4-1 Okubo, Shinjuku, Tokyo 169-8555, Japan}}
\newcommand{\YNU}{\affiliation{Department of Physics, Faculty of Engineering, Yokohama National University, Yokohama, Kanagawa 240-8501, Japan}}

\ICRR
\SNU
\IBS
\ISEE
\IPMU
\KMI
\Kobe
\KRISS
\Miyagi
\Nihon
\Tokai
\Tokushima
\Tsinghua
\Waseda
\YNU

\author{K.~Abe}\ICRR \IPMU
\author{K.~Hiraide}\ICRR \IPMU
\author{N.~Kato}\ICRR
\author{S.~Moriyama}\ICRR \IPMU
\author{M.~Nakahata}\ICRR \IPMU
\author{K.~Sato}\ICRR
\author{H.~Sekiya}\ICRR \IPMU
\author{T.~Suzuki}\ICRR
\author{Y.~Suzuki}\ICRR
\author{A.~Takeda}\ICRR \IPMU
\author{B.~S.~Yang}\SNU
\author{N.~Y.~Kim}\IBS
\author{Y.~D.~Kim}\IBS
\author{Y.~H.~Kim}\IBS \KRISS
\author{Y.~Itow}\ISEE \KMI
\author{K.~Martens}\IPMU
\author{A.~Mason}\IPMU
\author{M.~Yamashita}\IPMU
\author{K.~Miuchi}\Kobe
\author{Y.~Takeuchi}\Kobe \IPMU
\author{K.~B.~Lee} \KRISS
\author{M.~K.~Lee} \KRISS
\author{Y.~Fukuda} \Miyagi
\author{H.~Ogawa} \Nihon 
\author{K.~Ichimura}\RCNS 
\author{Y.~Kishimoto}\RCNS \IPMU
\author{K.~Nishijima}\Tokai
\author{K.~Fushimi}\Tokushima
\author{B.~D.~Xu} \Tsinghua \IPMU
\author{K.~Kobayashi} \Waseda 
\author{S.~Nakamura} \YNU

%Collaboration name if desired (requires use of superscriptaddress
%option in \documentclass). \noaffiliation is required (may also be
%used with the \author command).
%\collaboration can be followed by \email, \homepage, \thanks as well.
\collaboration{XMASS Collaboration}
\email{xmass.publications19@km.icrr.u-tokyo.ac.jp}
\noaffiliation

\begin{abstract}
Various WIMP dark matter searches using the full data set of XMASS-I, a single-phase liquid xenon detector, are reported in this paper.
Stable XMASS-I data taking accumulated a total livetime of 1590.9 days between November 20, 2013 and February 1, 2019 with an analysis threshold of ${\rm 1.0\,keV_{ee}}$. In the latter half of data taking a lower analysis threshold of ${\rm 0.5\,keV_{ee}}$ was also available through a new low threshold trigger. 
Searching for a WIMP signal in the detector's 97~kg fiducial volume %, which contained 97~kg of liquid xenon, 
yielded a limit on the WIMP-nucleon scattering cross section of 
${\rm 1.4\times 10^{-44}\, cm^{2}}$ for a ${\rm 60\,GeV/c^{2}}$ WIMP %${\rm 2.2\,GeV/c^{2}}$ 
at the 90$\%$ confidence level.
%Searching for an annual modulation signature in the full target volume, which contained 832~kg of liquid xenon, yielded a flux limit of  
%${\rm 2.3\times 10^{-42}\, cm^{2}}$ for a ${\rm 8\,GeV/c^{2}}$ WIMP from not seeing a nuclear recoil signal, ${\rm 1.4\times 10^{-35}\, cm^{2}}$ for a ${\rm 0.5\,GeV/c^{2}}$ WIMP using the Migdal effect, and  ${\rm 1.1\times 10^{-33}\, cm^{2}}$ for a ${\rm 0.5\,GeV/c^{2}}$ WIMP looking for bremsstrahlung from nuclear recoil. 
We also searched for WIMP induced annual modulation signatures in the detector's whole target volume, containing 832~kg of liquid xenon. For nuclear recoils of a ${\rm 8\,GeV/c^{2}}$ WIMP this analysis yielded a 90\% CL cross section limit of ${\rm 2.3\times 10^{-42}\, cm^{2}}$. Annual modulation signatures from the Migdal effect and Bremsstrahlung at a WIMP mass of ${\rm 0.5\, GeV/c^{2}}$ were evaluated and lead to 90\% CL cross section limits of ${\rm 1.4\times 10^{-35}\, cm^{2}}$ and ${\rm 1.1\times 10^{-33}\, cm^{2}}$ % were obtained for the Migdal effect signal and with bremsstrahlung, 
respectively.
\end{abstract}

%\date{October 2021}

\maketitle
%https://www.overleaf.com/project/617165d3604f3ae2645ac9ed

\section{Introduction}
\label{sec:Introduction}

Cosmological and astrophysical observations require the existence of dark 
matter (DM), and hypothetical DM particles provide a %consistent
compelling explanation for the
observed phenomena~\cite{Faber,Beringer}.
However, the properties of these hypothetical DM particles are
unknown, and none of the particles in our standard model of particle physics is a 
valid candidate. % for particulate DM.
%The direct detection of a DM particle interacting
%with normal matter was the primary goal of the XMASS-I %experiment~\cite{XMASS_det}.

One well-motivated DM particle candidate that might be detectable in direct 
detection experiments 
%like XMASS-I
is the weakly interacting massive particle 
(WIMP)~\cite{Goodman}.
% If WIMPs were carriers of
As WIMPs are postulated to share some weak force other than gravity % that they would share
with normal matter, this force would mediate interactions with materials that could be used as targets in a detector.
%the target material in our detector that results
These interactions %would then 
could result in the detectable recoil of individual target nuclei from such
WIMP interactions~\cite{Gondolo}. 
Many experiments are looking for various WIMP 
interaction signatures~\cite{XENON1T_1ty,XENON1T_Migdal,LUX,PandaX-4T,DARKSide,DEAP3600}. 
%~\cite{LZ2022,XENON1T_1ty,XENON1T_Migdal,LUX,PandaX-4T,DARKSide,DEAP3600}. 

The XMASS-I was one of these experiments, and the primary goal of this XMASS-I experiment was to detect directly a DM particle interacting with normal matter~\cite{XMASS_det}.
The XMASS collaboration previously published WIMP search results~\cite{XMASS_LowMassWIMP,XMASS_inelastic,XMASS_Modulation,XMASS_bosonic,XMASS_FV,XMASS_Modulation2018,XMASS_Modulation2019,XMASS_inelastic2019} 
from its unique large volume, single-phase design for liquid xenon (LXe) detector using only scintillation signals in its 832\,kg liquid xenon target mass.
% In addition to that, their
Beyond this primary goal, versatility has also allowed detectors like XMASS-I to address a much wider range of physics topics, making it possible for the XMASS collaboration to publish results on dark photons, axions, and axion-like particles~\cite{XMASS_AXION2013,XMASS_KK,XMASS_hiddenphoton}, double electron capture~\cite{XMASS_DEC,XMASS_DEC2018}, neutrinos~\cite{XMASS_SN,XMASS_exotic}, 
and coincidences with gravitational waves~\cite{XMASS_GW}.

This paper is based on data from a full XMASS-I exposure collected underground at the Kamioka Observatory in Japan during five years of stable data taking.
It presents various rare-event searches using nuclear recoil (NR) and electron recoil (ER) signatures. %channels. 
Section~\ref{sec:Detector} provides a brief description of the detector and Sec.~\ref{Sec:Calib} of its calibration. 
% Event selections 
Common event selection steps for the analyses in this paper %'s analyses 
are detailed in Sec.~\ref{sec:Dataset}.
Section~\ref{Sec:FV} presents a WIMP search %by fitting 
which directly fits the energy spectrum in a limited detector volume, 
and Sec.~\ref{Sec:modulation} extends our modulation searches to the full XMASS I data set. 
Conclusions are %given
offered in Sec.~\ref{Sec:Conclusions}.

\section{Detector}
\label{sec:Detector}
The XMASS-I detector \cite{XMASS_det}, underground at the Kamioka Observatory in Japan with an overburden equivalent to 2,700 meters of water, was a single-phase LXe detector built to detect DM particles. 
Its almost spherical inner detector (ID) volume contained 832 kg of LXe, had a radius of $\sim$40~cm, and was viewed by 630 hexagonal %photomultiplier tubes (PMTs) 
“HAMAMATSU R10789-11” and 12 round “HAMAMATSU R10789-11MOD" photomultiplier tubes (PMTs)~\cite{XMASS_R10789}.
At the LXe scintillation wavelength of 175 nm the quantum efficiency of these PMTs was 30\% on average, and their photocathodes %of these PMTs 
covered 62.4\% of the inner surface of the active LXe volume in the detector.
The PMTs were held in 60 copper triangles % which produced the actual
% made of oxygen-free high conductivity (OFHC) copper that
%60 oxygen-free high conductivity (OFHC) copper triangles that %defined
which gave the detector's inner surface its actual pentakis-dodecahedral shape. % of the detector's inner surface.
During the commissioning phase we found that the aluminum seal between the PMTs' quartz windows and their metal bodies contained the upstream portions of the $^{238}$U and $^{210}$Pb decay chains. To mitigate the effect of this, the region where the PMT windows meet their PMTs' metal bodies were 
% sealed from the outside for each PMT with 
covered with a copper ring during the ensuing detector refurbishment. 
The gaps between neighboring PMTs' copper rings were furthermore covered with thin copper triangular plates with cutouts for the PMTs' photocathode areas for each of the triangles in the pentakis-dodecahedral inner surface. 
%for each of the triangles in the pentakis-dodecahedral inner surface. 
%The gaps between neighboring PMTs' copper rings were furthermore covered with thin copper plates, one triangluar plate with cutouts for the PMTs' photocathode areas for each of the Cu-triangles in the pentakis-dodecahedral inner surface. 
To limit radiogenic background in the LXe target all these structural elements, including both the inner and outer cryostat themselves, were fabricated from oxygen-free high conductivity (OFHC) copper. 

The cold ID inside this cryostat was shielded from the surrounding rock's gamma and neutron emissions by a cylindrical water shield of 11~m height and 10~m diameter. 
This water shield, which also served as an active muon veto, contained ultra pure water and was instrumented with 72 Hamamatsu R3600 20-inch PMTs, 
%; hence it is
and is referred to as the outer detector (OD). 

To reduce the detector's exposure to ambient $\rm {}^{222}Rn$, the whole underground experimental hall in which XMASS-I was housed was lined with radon retarding material and continually flushed with fresh air forced in from above ground.  % outside. %Altogether XMASS Rn mitigation measures limited the Rn concentration in the OD to 150 mBq/m$^3$.
To the top air layer of the OD tank, Rn removed air was supplied.
The Rn concentration in the water of the OD was continuously monitored from April 16, 2014 and found to be less than 150~mBq/m$^{3}$.

Readout of the XMASS-I data acquisition (DAQ) system was triggered by a multiplicity trigger based on Analog-Timing-Modules inherited from Super-Kamiokande ~\cite{SK_ATM}.  When a PMT signal surpassed a set threshold, these modules produced a 200~ns wide standard analog output signal for each threshold crossing on the corresponding channel. This single-channel PMT signal threshold was set to 0.2~photoelectron (PE) equivalent for the ID and 0.4~PE equivalent for the OD. Simple analog summation of the resulting 200~ns flat-top analog signals of all channels in linear fan-in/fan-out units then allowed triggering event readout by setting the appropriate multiplicity threshold on the resulting analog sum. This summation was done separately for both ID and OD PMTs, resulting in independent ID and OD triggers.

During all of XMASS-I data taking, which started in 2013 after the detector refurbishment, ID DAQ readout was thus triggered when four or more ID PMTs passed their single-channel threshold within 200 ns of each other, and OD DAQ readout was initiated when at least eight OD PMTs crossed their thresholds by a parallel analog summation of OD PMT threshold crossings. 
Flash analog to digital converters (FADCs), CAEN V1751, recorded waveforms from 1~$\mu$s before to 9~$\mu$s after an % ATM  system had issued a detector readout signal for each PMT that would cross its single channel trigger threshold within the readout window. 
ID trigger was issued for each ID PMT which crossed its single-channel threshold within this latter time window (0-9 $\mu$s).
From 2015 December on an independent low-threshold trigger was added %independently 
for the ID, requiring only % which only required
three PMTs to cross their threshold within 200~ns of each other to initiate an ID DAQ readout also for such low-threshold events. 
One-pulse-per-second (1PPS) signals from a global positioning system (GPS) receiver were
used to trigger readouts independent of PMT signal coincidences, adding time stamps to the data stream.
These 1PPS GPS signals also flashed a light-emitting diode (LED) in the ID's inner surface, which added single PE (SPE) signals that were used for PMT gain monitoring.

\section{Calibration}
\label{Sec:Calib}

\subsection{PMT calibration}

\begin{figure}[htbp]
  \begin{center}
    \includegraphics[width=0.5\textwidth]{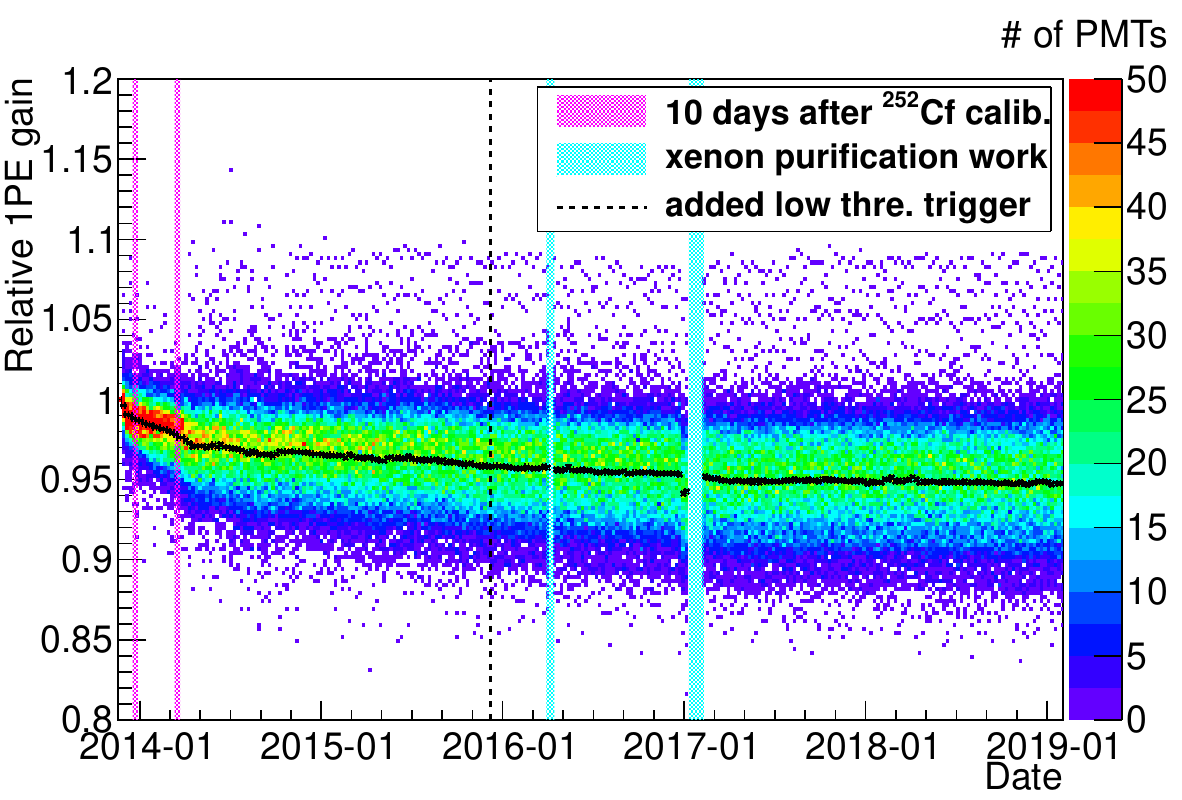} 
  \end{center}
  \caption{Time evolution of ID SPE gains relative to that in the first week of data taking for each PMT.
  The black points represent relative gain averaged over all ID PMTs each week.
  %The colored points in the underlying scatter plot show that of  individual PMTs.
  The colored points in the underlying scatter plot show the individual PMTs; the color scale represents the number of PMTs in the 0.0025 gain binning each averaged over that same week.
  The black dashed vertical line indicates the start of low-threshold data acquisition.
The cyan and magenta bands indicate periods when data acquisition was interrupted by the Xe purification work and neutron activation after neutron source calibration. 
  }
  \label{fig:1pe-gain-history}
\end{figure}

As indicated above, ID PMT gains were monitored with a blue LED embedded at the inner surface of the ID, triggered by a 1PPS GPS signal. 
The LED's intensity was first adjusted for SPE-compatible occupancy in LED-associated events.
Then, the gain of each ID PMT was calculated from the SPE event mean charge in the LED data, accumulated 
%averaged 
over one week.
Figure~\ref{fig:1pe-gain-history} shows the time evolution of all ID SPE gains relative to the first week's gain. 
% for each PMT.
A gradual decrease in this ID PMT gain was observed over the entire data-taking period.
A gain drop of about 1\%  was observed after conducting the neutron calibration in December 2016 using the Y/Be source which generated a high rate of bright light events. That drop was recovered though after the Xe purification work.
All PMT's gain were equalized to $10^{7}$ at the onset.
The observed gain evolution in each PMT was then corrected to convert the detected charge to the number of PEs.

\begin{figure}[htbp]
  \begin{center}
    \includegraphics[width=0.5\textwidth]{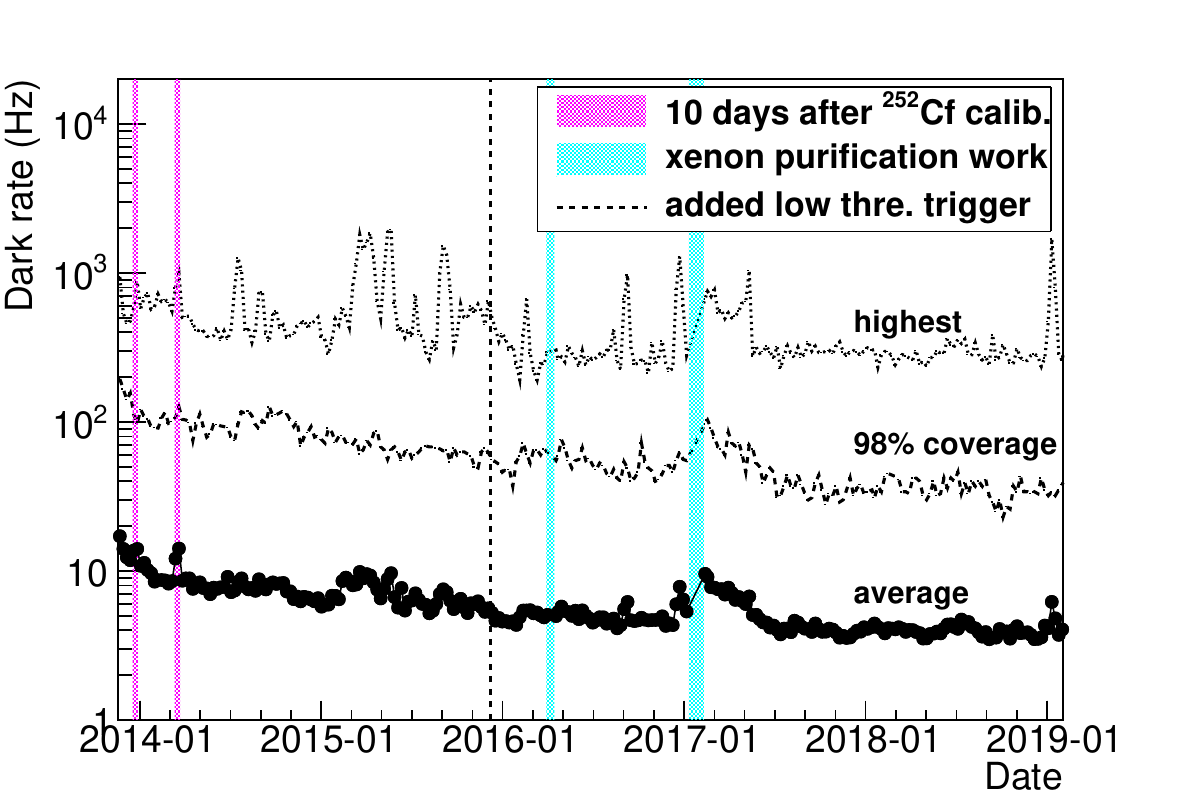}
  \end{center}
  \caption{Time evolution of the weekly dark rate averaged over all PMTs (points),
  the ``98\% coverage'' rate where 98\% of the PMTs have a smaller dark rate than this value (dashed), and the highest rate of any single PMT (dotted).
  }
  \label{fig:single-rate-history}
\end{figure}

The 1PPS data also served to monitor the ID PMTs' dark rates by counting the number of hits in a 1~$\mu$s window before the LED flash.
Figure~\ref{fig:single-rate-history} shows the time evolution of a weekly dark rate averaged over all ID
PMTs, together with the ``98\% coverage'' rate where 98\% of the PMTs have a smaller dark rate than this value, and the highest rate of any single PMT. %; again the data were averaged over one week.
% The average single rate in each PMT was approximately 15~Hz
The average dark rate in all ID PMTs was 15~Hz
% at earlier periods 
in the beginning and had decreased to about 5~Hz by the end of data taking.
This decrease in the dark rate eventually allowed us to lower the data taking and analysis thresholds.
%during the operation.
%over time.

\begin{figure}[htbp]
  \begin{center}
    \includegraphics[width=0.5\textwidth]{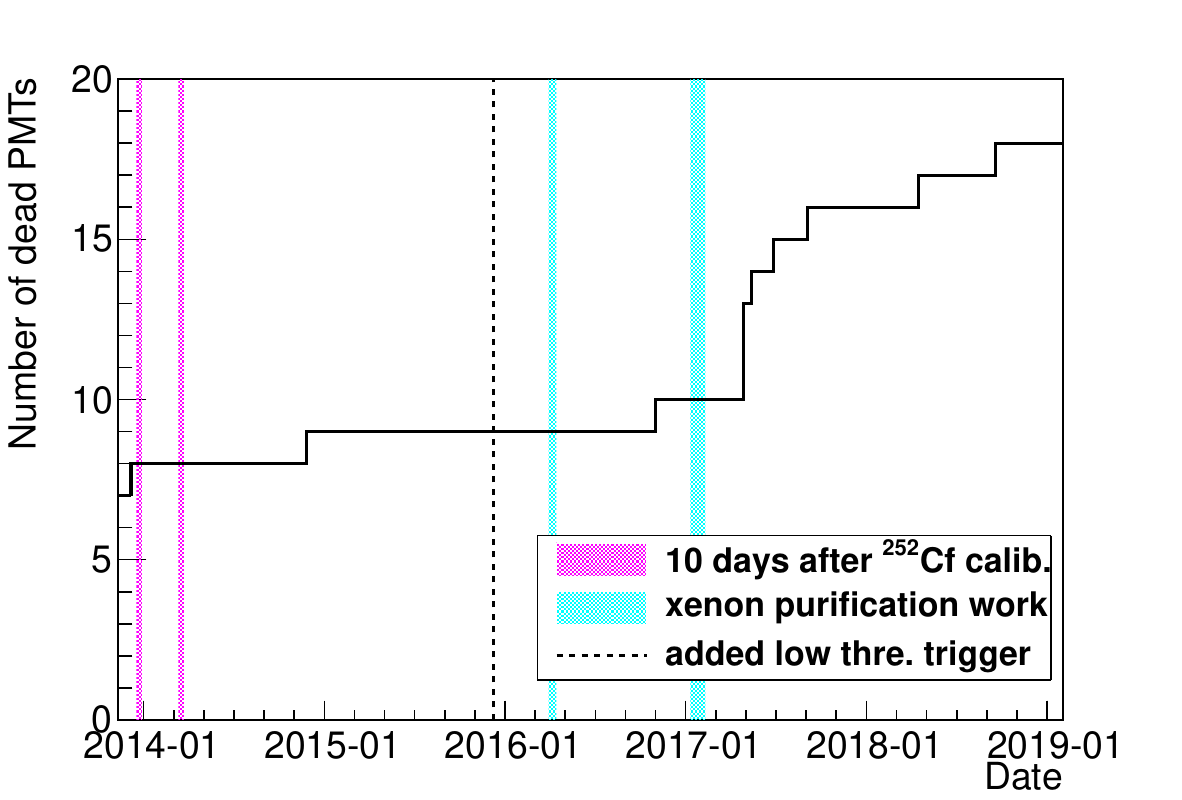}
  \end{center}
  \caption{Number of dead ID PMTs over time.}
   %Time evolution of the number of dead PMTs in the ID.}
  \label{fig:dead-pmt}
\end{figure}

When a PMT malfunction occurred, its PMT was turned off and the data from it was removed from the analysis process. This was because high rate noise often had an uncontrollable negative impact on data quality. The number of nonoperational PMTs in the ID, hereafter referred to as dead PMTs,
is shown in Fig.~\ref{fig:dead-pmt} and rose from 7 to 18 over the total data-taking period. 
Before the large increase around April 2017, the detector had temperature cycle 
related to the 2nd xenon purification work, including warming up to room temperature.
%, as shown in Fig.~\ref{fig:dead-pmt}.
%Most of these PMTs had to be turned off because their respective hit rate increase threatened to compromise the DAQ. 
% Data from the dead PMTs were discarded for that failure happened run and all subsequent runs.
%When a PMT had to be declared dead, data it provided in the run in which it failed and any subsequent runs were discarded from our analyses.

\begin{figure}[hbtp]
\centering
\includegraphics[width=0.52\textwidth]{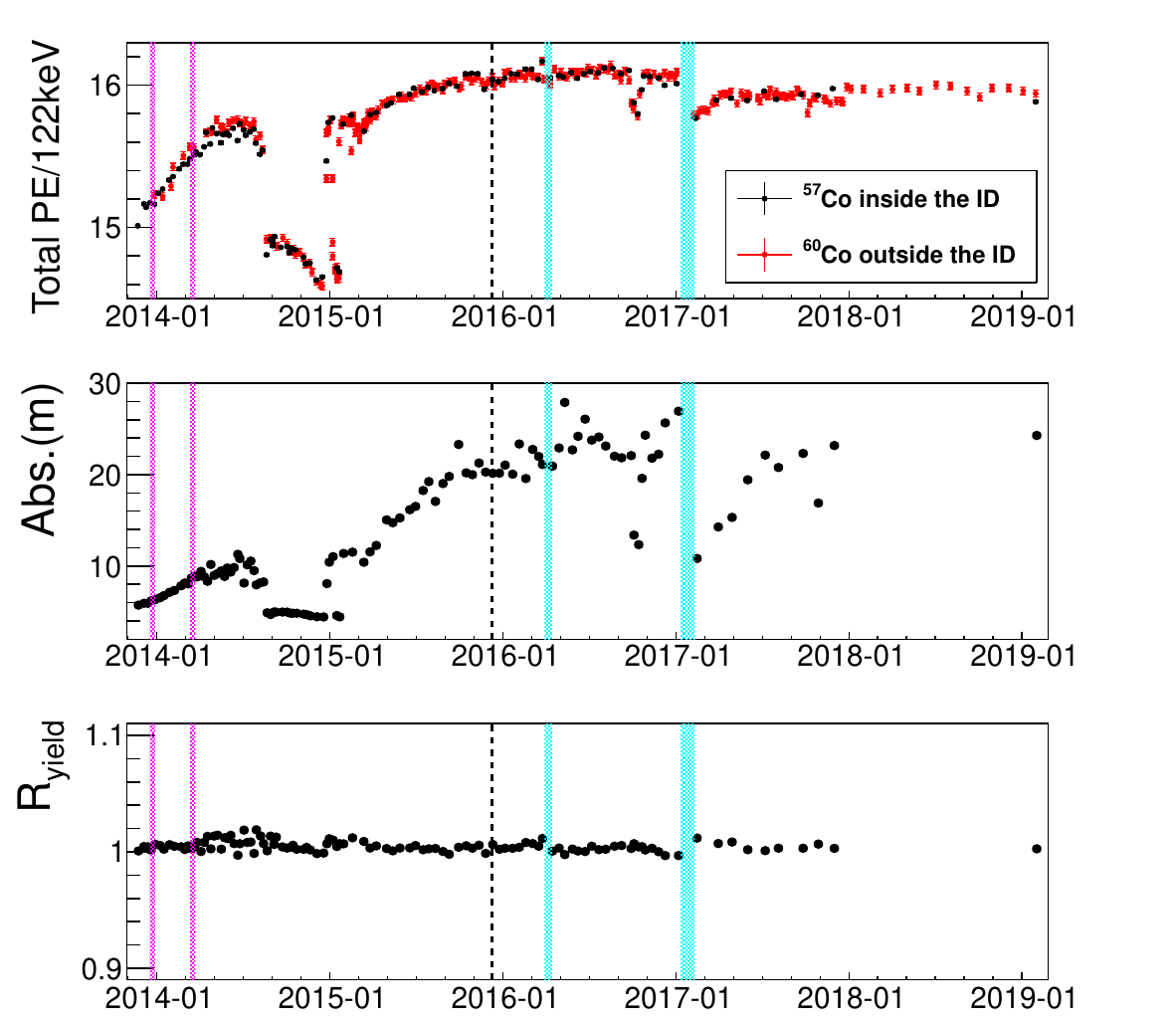}
\caption{The top panel shows the observed total PE per 122~keV gamma rays,
the black markers show $^{57}$Co measurements made with the source inside the ID, and the red ones refer to data taken with a $^{60}$Co irradiating the LXe target from a fixed position in the OD.
The absorption length for LXe scintillation light is shown in the middle panel.
The bottom panel shows the scintillation light yield ($R_{yield}$) relative to the value at the start, without taking into account the light propagation or PMT effects.
The absorption length and the light yield were derived from regular $^{57}$Co calibrations. 
% regularly with either a $^{57}$Co (black markers) inside the detector or a $^{60}$Co (red markers) outside the ID. 
  }
\label{fig:stability}
\end{figure}

\subsection{LXe scintillation light: yield and propagation}
% PE yield and propagation} % Xe parameter} % 

% Figure \ref{fig:stability} shows the time variation of the detector's PE yield and absorption length for LXe scintillation light and also the %relative
% intrinsic scintillation light yield.

The PE yield in the XMASS-I detector was tracked by inserting a $^{57}$Co source into the ID every one or two weeks.
From these $^{57}$Co calibration data~\cite{XMASS_cal}, taken at nine different positions along the vertical z-axis through the center of the detector from $z$ = $-$40~cm to  $z$ = +40~cm, 
the absorption and scattering lengths for scintillation light as well as the light yield of the LXe scintillator were inferred by matching the PE hit patterns in data with those of Monte Carlo (MC) simulation.
The probability of the simultaneous emission of two PEs for a single LXe scintillation photon striking the photocathode of our ID PMTs~\cite{dpe1,dpe2} was also properly taken into account in all our simulations.

Another way of tracking these LXe scintillation light properties was by temporarily placing a $^{60}$Co source at the same position of the OD outside of the ID's cryostat.
A good linear correlation between the PE yield of $^{57}$Co and that of $^{60}$Co was 
then confirmed by 
comparing both sets of data taken on the same day during normal operation periods. 
This comparison allowed us to express the $^{60}$Co data in PE per 122\,keV gamma ray, the natural unit for the standard $^{57}$Co calibration.
Once the detector was stable towards the end of 2017, we suspended the more disruptive $^{57}$Co calibration for which the source had to be inserted into the ID.

In Fig.~\ref{fig:stability} the top panel shows the resulting yield measurements and their time evolution throughout XMASS-I data taking; the black markers show $^{57}$Co measurements made with a source at $z$ = $-$30~cm inside the ID, and the red ones refer to the $^{60}$Co measurements, with their source deployed in the OD. 
The variations in the absorption length and the relative scintillation light yield $R_{yield}$ 
of the LXe scintillator extracted from $^{57}$Co calibration are 
shown in the middle and the bottom panels, 
%in Fig.~\ref{fig:stability}
respectively.
As can be seen from the figure, $R_{yield}$ was close to constant, varying within 1--2\% at most. 
The scattering length also remained stable at around 52~cm.
In our MC, an OFHC copper reflectance of 0.25+-0.05 as specular reflection was used for LXe scintillation light for the entire period. We evaluated this reflectivity using the events in the ~92 keV gamma-ray peak from the progeny of $^{238}$U ($^{234}$Th) in the PMT aluminum seal by comparing the data and MC.
The observed changes in PE yield can be explained by corresponding variations in the absorption length. 
The abrupt changes around August 2014 and December 2014 were due to power outages and subsequent work undertaken to remove impurities released into the LXe during the outages. As the top panel in Fig.~\ref{fig:stability} shows, the accompanying excursions in the ID's PE yield were largely driven by such absorption length changes~\cite{XMASS_Modulation2018}.
% These variation in PE yield can be explained by changes of this absorption length~\cite{XMASS_Modulation2018}. 
% To suppress absorption gas evaporating from the LXe surface in the ID was continually purified by circulation through a hot zirconium getter after March 2015.
From March 2015, the Xe gas evaporating from the liquid in the ID was routinely passed through a hot getter (API-Getter II, API) before being liquefied again and returning to the ID.
The xenon purification work at the beginning of 2017 was a distillation to further reduce the Ar level in LXe. The decrease in PE yield and absorption length after this distillation is thought to be due to impurities (water) trapping on the refrigerator unit's cold finger, which were released when the operation status of the one unit was changed and the unit warmed up.

\subsection{Energy scale}

Since the scintillation efficiency in LXe depended on the density of the energy deposit along a particle track, the estimation of energy deposited in the detector from the observed scintillation light depended on the particle that deposited the energy.
Two energy deposits are of particular interest in analyzing LXe detector data: that from ER of a single electron and that from NR of a whole Xe nucleus.
%A certain scintillation signal in the detector will thus either be interpreted as stemming from an ER or NR.
%electron depositing an amount of energy measured in 
%Interpreting the scintillation light measured, 
%as deposited by an electron 
We denoted 
the amount of energy an electron would have to deposit to produce an observed amount of scintillation light as keV$_{\rm{ee}}$, and the energy a recoiling Xe nucleus would have to deposit as keV$_{\rm{nr}}$. 
The corresponding ER and NR energy scales used in this paper's analyses are the same as the ones used in our previously published sub-GeV DM analysis~\cite{XMASS_Modulation2019}.
%and the scintillation efficiency of NR signals was taken from Ref.~\cite{XENON2011}.
%In the case of electron-equivalent energy, the model from Doke et al. \cite{DOKE} corrected using a gamma-ray source is used for the the non-linearity of the light yield (scintillation efficiency).
\begin{figure}[htbp]
\centering
    \includegraphics[width=0.45\textwidth]{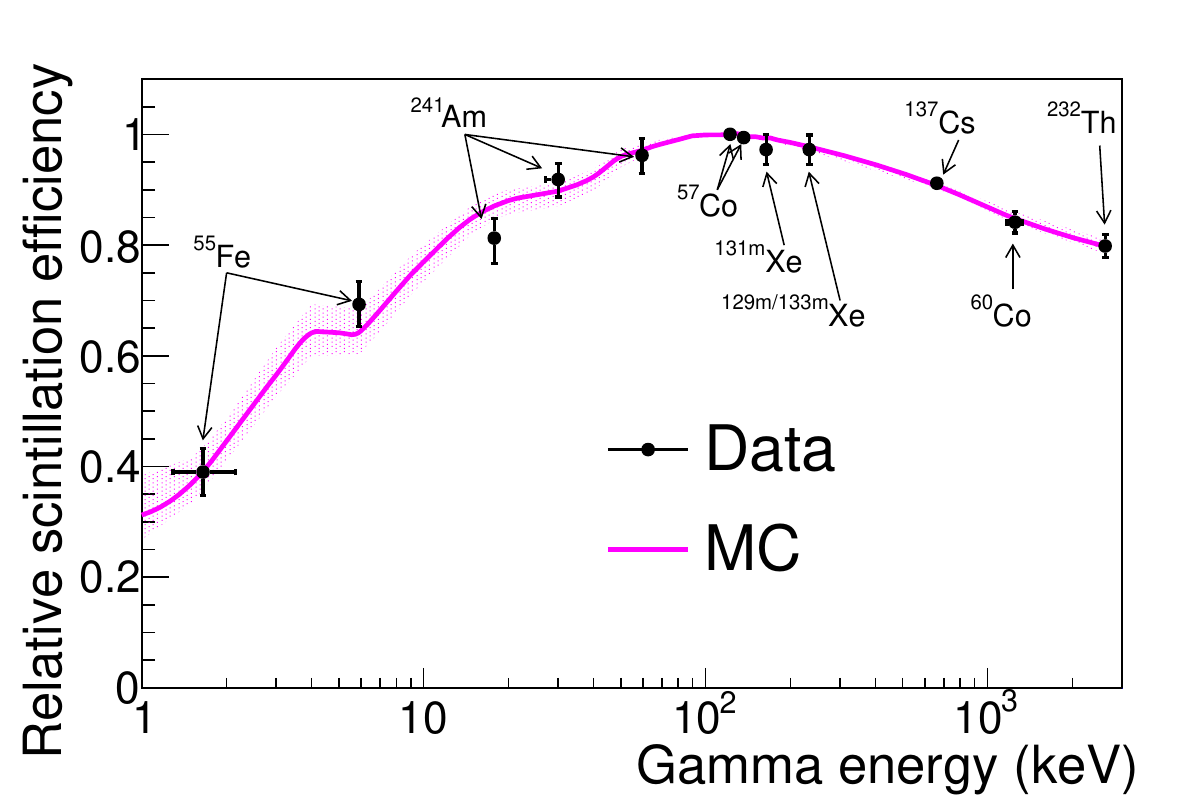}
  \caption{
  Relative scintillation efficiency as a function of gamma-ray energy.
  The efficiency at 122\,keV was set to 1.
  The black data points were measured 
  using gamma-ray sources in the detector, the magenta curve is what we implemented in our MC, and the band represents its one sigma error band.
  1.65 keV by $^{55}$Fe and 30.0 keV by $^{241}$Am are the escape X-ray peaks (escaped into the source container).
  }
  \label{fig:scint_eff}
\end{figure}

Figure \ref{fig:scint_eff} shows the relative scintillation efficiency 
of gamma lines observed in XMASS-I;
it is normalized to one at a $^{57}$Co calibration line of 122\,keV.
Data points and their error bars come from
various gamma lines observed in the spectra of the calibration sources used in our detector. 
The line represents how these measurements were interpolated in our MC, with the underlying band representing the variance used in the studies of our systematic uncertainties.
Among the sources, $^{55}$Fe, $^{241}$Am, $^{57}$Co were introduced into the ID along the z-axis, 
whereas $^{60}$Co and $^{232}$Th sources were applied to a specified position in the OD, just outside the cryostat. % are used outside the vacuum chamber.
We also used the xenon isotopes $^{131m}$Xe,$^{129m}$Xe, and $^{133m}$Xe produced %after
during neutron calibrations in our gamma-ray calibration.
The efficiency below 5.9\,keV was calibrated by the L-shell X-ray escape peaks,  measured during calibration with an $^{55}$Fe source.
These escape peaks distributed energy in the 1.2--2 keV interval, with the weighted mean energies of these escape peaks being 1.65 keV and having an RMS of 0.43\,keV \cite{XMASS_Modulation2019}.

The electron-equivalent energy scale used % below
in this paper was constructed using the results of electron simulations based on the relative scintillation efficiency, as discussed above.

\section{The data set}
\label{sec:Dataset}

\subsection{Data-taking overview}
\label{sec:dataoverview}
\begin{figure}[htbp]
  \begin{center}
    \includegraphics[width=0.5\textwidth]{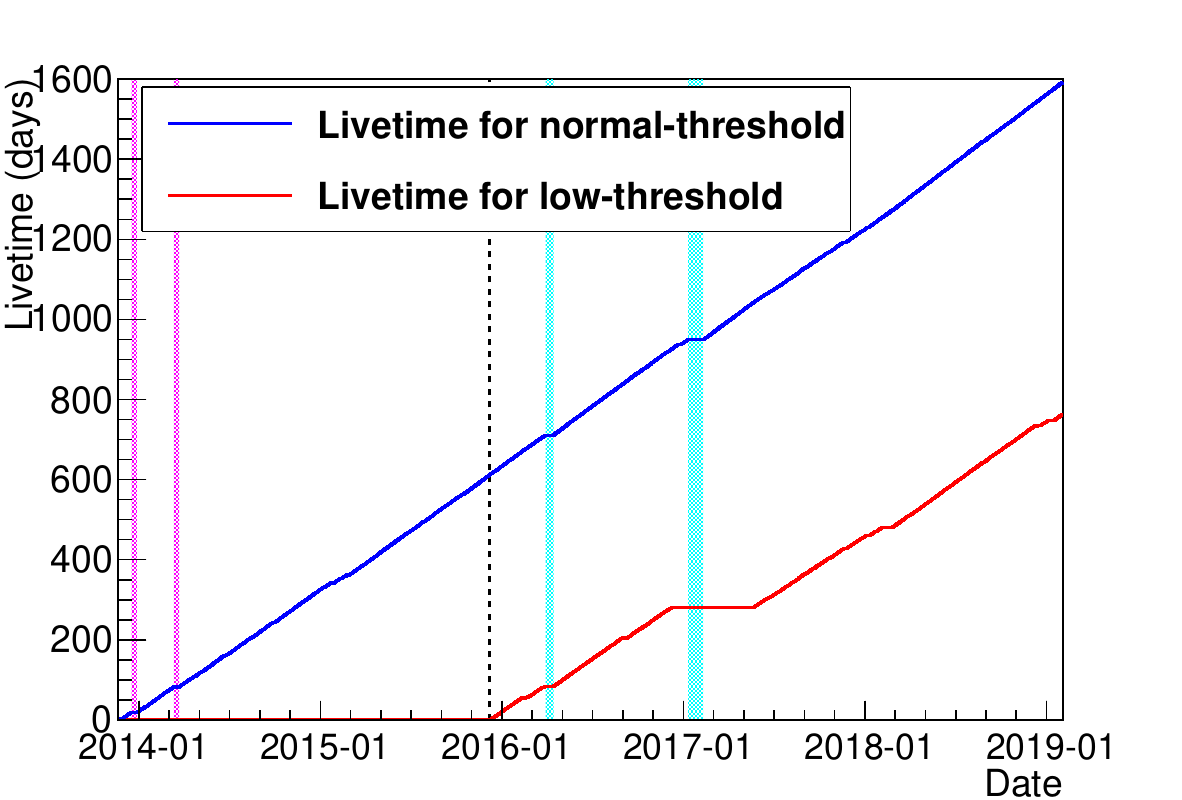}
  \end{center}
  \caption{% Accumulated detector livetime after standard cuts for the normal-threshold data
  Livetime accumulation for normal- 
  (blue) and low-threshold (red) data.
  The black dashed vertical line indicates the start of low-threshold data taking.
  The magenta bands represent 10-day periods after detector calibration with the $^{252}$Cf source.
  The cyan bands show periods of xenon purification work.}
  \label{fig:livetime}
\end{figure}

The data used in this analysis were collected between November 20, 2013 and February 1, 2019. Normal data taking was % artificially broken up into
organized in 24 hour "runs" for bookkeeping and data management unless there was a specific reason %not to.
to terminate a run earlier.
%organized in "runs", defined by DAQ start to stop, for bookkeeping and data management. 
Figure~\ref{fig:livetime} shows the XMASS-I livetime accumulation over time. Data taking was interrupted twice for a few weeks (cyan colored bands in the figure): the first time, all LXe were removed from the detector so that they could be passed through a getter (SA-MT15, SAES) upon re-insertion, which removed impurities inadvertently released from a warming cold head. The second interruption allowed a distillation campaign to remove argon (Details are given in Sec.~\ref{sec:BGest}). We twice used a $^{252}$Cf neutron source and had to wait for the neutron activation in the detector to abate, shown in the figure as magenta colored bands. $^{252}$Cf calibration data was used for the study of the NR scintillation decay time in LXe~\cite{XMASS_neutron}.
Regular ID % light yield 
calibrations were another source of dead time.

After accounting for dead time and analysis specific run selection (see below) the total collected XMASS-I livetime was 1590.9 days. Additional low threshold data taking started on December 8, 2015
with its own run selection criteria, resulting in 768.8 days of XMASS-I livetime with low threshold data.

\subsection{Run selection}
\label{Sec:Data:run}
\begin{figure}[htbp]
  \begin{center}
    \includegraphics[width=0.48\textwidth]{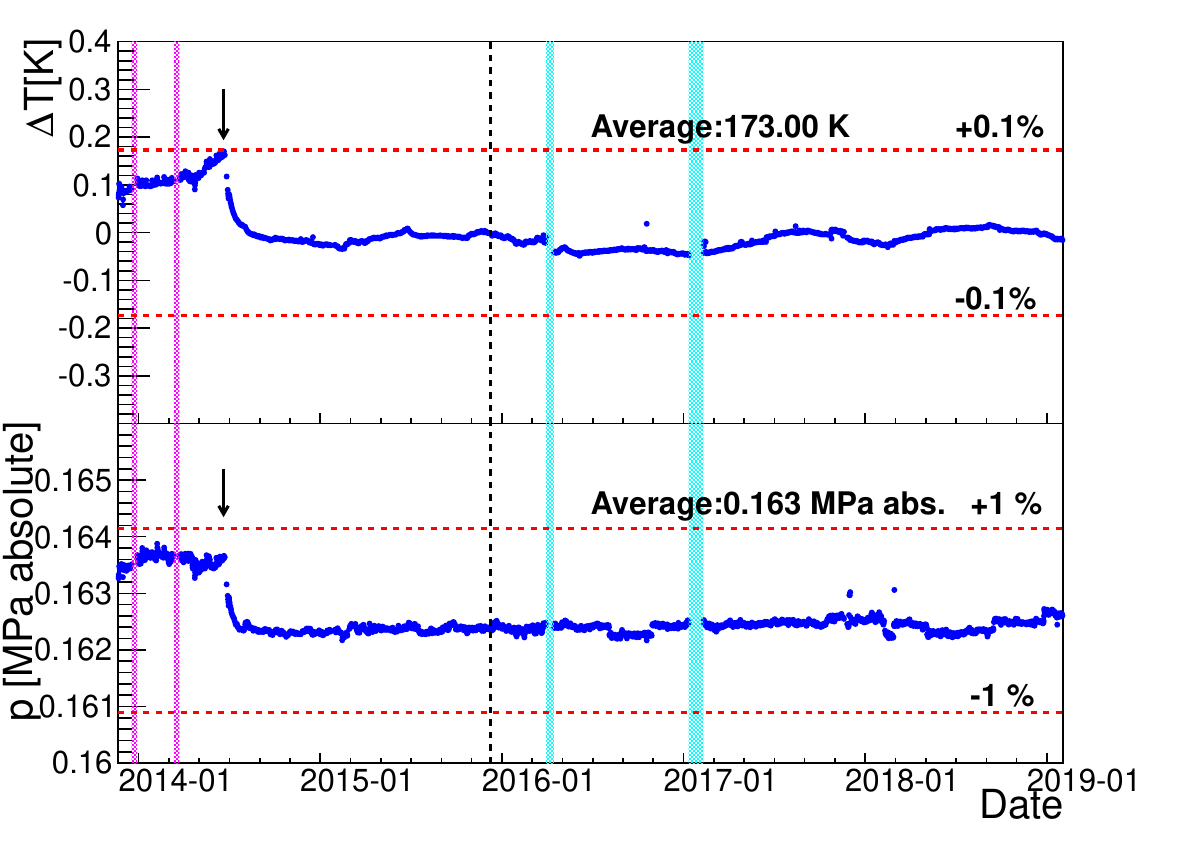}
  \end{center}
  \caption{Time evolution of the temperature in and gas pressure above the LXe in the ID.
For the temperature, its difference from the nominal temperature (see the text) is shown. 
The drop at the arrow was due to a change of reference sensor for controlling the cold-head that re-liquefied evaporated Xe from the target mass.}
 % refrigerator. (... in the lunch room???}
  \label{fig:temperature-pressure}
\end{figure}

Runs were considered for the physics analyses presented here if they lasted at least one hour and had no DAQ problems.
Figure \ref{fig:temperature-pressure} shows the stability of the %detector temperature measured in the LXe
LXe temperature in the ID and the pressure above the ID's liquid surface throughout all of XMASS-I data taking. % that whole time.
The nominal temperature and pressure of the detector were 173.0~K and 0.163~MPa, respectively.
Data runs were analyzed only if % in our analyses only if
their temperature and pressure were within 
% We selected runs during which the temperature and pressure are stable within 
$\pm$0.05~K and $\pm$0.5~kPa % from each reference points, respectively
% The the reference values are changed after inner calibrations and power outages, etc.
of those of their respective ID calibration runs.
% After taking data with both lower and higher pressure in the ID, these criteria were set to keep the PE yield change less than 0.1\%, negligibly small.
This ensured that changes in scintillation light yield were $<$ 0.1~\%, 
which was verified in a study where LXe temperature and pressure were changed and their impact on PE yield checked.  
In this study during detector commissioning before the physics run, $^{57}$Co source calibration was performed under different liquid xenon pressures in the XMASS detector. A 6.6$\pm$0.4~\% change in light intensity was observed for a pressure change from 0.129 to 0.231~MPa with a temperature change of about 12~K.
The largest temperature drop of about 0.2~K happened on Jun.~23, 2014, 
%when the sensor providing the reference temperature for the ID temperature control loop was changed; 
when the sensor providing the reference temperature for the ID temperature control loop was changed from one installed directly on the copper cold finger attached to the refrigerator to one in the pipe returning the liquefied xenon to the detector;
no impact on the PE yield is evident in the top panel of Fig.~\ref{fig:stability}. 
% This drop was due to a change of the reference sensor for the feedback loop that controls the detector temperature
% Several other drops or peaks are due to power outages, inner source calibrations, etc.
% Runs with data acquisition problems or with a duration less than 1 hour were also removed from the data set. 

%Next we looked at ID and OD trigger rate changes within each run, we averaged the trigger rate in every 10 minutes and that should within 5~sigma from that run’s average rate.
Next the ID and OD trigger rates in each run were averaged over 10 minute intervals: a run would remain in the data set if none of the 10 minute averages were more than 5~sigma from the run average. 
%\textcolor{red}{What does sigma mean here: the sigma of the distribution of 10 min averages for each individual PMT or sqrt(average for the run)? }=> (Ogawa, 220912) the sigma of the distribution of 10 min averages for "trigger rate". Upon sentence say it.
We also eliminated runs in which more than 20 triggers for ID and 5 triggers for OD are issued in any one second % to assure DAQ system stability. 
as DAQ problems may occur beyond these limits. 
To avoid effects from neutron activation in the detector, 
runs within 10 days after a $^{252}$Cf calibration were also excluded from the data sets. 
% to avoid the effect of xenon activation by neutrons.

%\subsection{Event selection}
\subsection{Standard cuts applied to all ID events}

A basic ID event selection, referred to as the standard cut,
% was applied before proceeding with physics analysis.
precedes all XMASS-I physics analyses.
In the following a threshold crossing in a PMT, which for the ID PMTs leads to a readout of its waveform digitizer if it belongs to a triggered event, will be referred to as a hit, registered on the respective PMT at the time of the threshold crossing.

First, we require that the ID event under scrutiny is not associated with an OD event, that no concurrent OD event (triggered by $\geq$ 8 hits in the OD within 200~ns window) exists.

\begin{figure}[htbp]
  \begin{center}
    \includegraphics[width=0.5\textwidth]{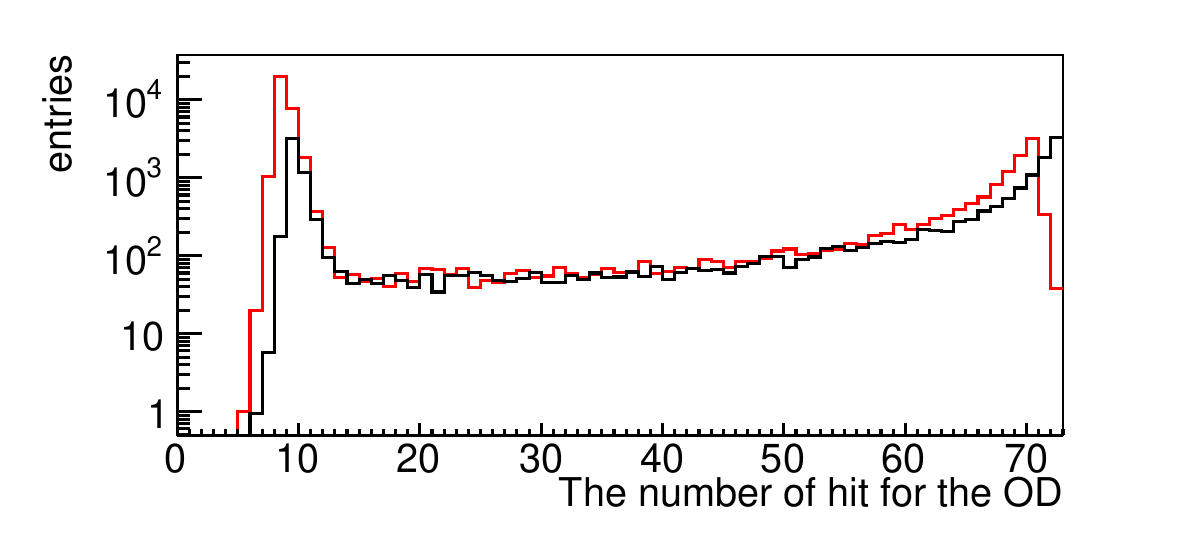}
  \end{center}
  \caption{The number of hit distributions for the OD in runs at the beginning (black) and the end (red) of the data-taking period.} %(bottom) The time variation of the OD event rate with the number of OD hits more than 20. Red dot line indicates the average rate for whole data-taking period.}
  \label{fig:OD}
\end{figure}

Figure~\ref{fig:OD} shows the number of hit distributions for the OD in two 24 hours runs, one at near the beginning (Jan. 2014) and the other at the end (Jan. 2019) of the data-taking period.
%Figure~\ref{fig:OD} (top) shows the number of hit distributions for the OD in runs at the beginning and the end of the data-taking period. %The OD data was taken if we had more than eight hits in the 200 ns window. 
The peak at the threshold, around ten, is mainly due to external radiation and electronic noise.
Over the whole data-taking period the OD event rate increased from 0.3 Hz to 0.7 Hz for $\geq$ 8 OD PMT hits. 
The possible reason for this was an increase in the electronic noise, which is due to one of the PMTs becoming noisy and one new DAQ fan module. The impact of this increase in OD event rate is negligible for the analysis in this paper.
%The effect of this increase in noise on the livetime is negligible.
The shift of the peak position around 72, shown in Fig.~\ref{fig:OD} is due to two dead OD PMTs during the data taking. To identify muon events, we required the number of OD hits more than 20. %Figure~\ref{fig:OD} (bottom) shows the time variation of the OD event rate with the number of OD hits more than 20. 
The observed muon event rate is approximately 0.17 counts per second for $\geq$ 20 OD PMT hits, which is stable throughout XMASS-I data taking and consistent with the estimated muon rate at the XMASS-I experimental site based on the muon rate at Super-Kamiokande~\cite{SK_PCRflux}.

Afterpulses are likely to occur in ID PMTs following a high energy event in the ID, and very often trigger a new ID readout all by themselves.
To avoid such events consisting mainly of afterpulses, only ID events for which the trigger time difference to the preceding ID event is longer than 10~ms are retained for analysis.
Counting 1PPS events affected by this cut we found that, averaged over the whole data taking period, this cut introduced 3\% of additional deadtime. 
% Furthermore, a cut on events with a standard deviation of hit timings greater than 100~ns is used to remove noise events caused by afterpulses.
Afterpulse events that do occur after even longer delays are cut by checking the spread of hit timings in the event: events with a standard deviation of hit timings greater than 100~ns are also discarded as afterpulse events. 
Decays of $^{40}$K in ID PMT's photocathodes often emit Cherenkov radiation in the PMT's quartz window. 
Events with more than 60\% of their PMT hits registered in the first 20~ns of the event are discarded as
Cherenkov events~\cite{XMASS_LowMassWIMP}. 

\begin{figure}[htbp]
  \begin{center}
    \includegraphics[width=0.5\textwidth]{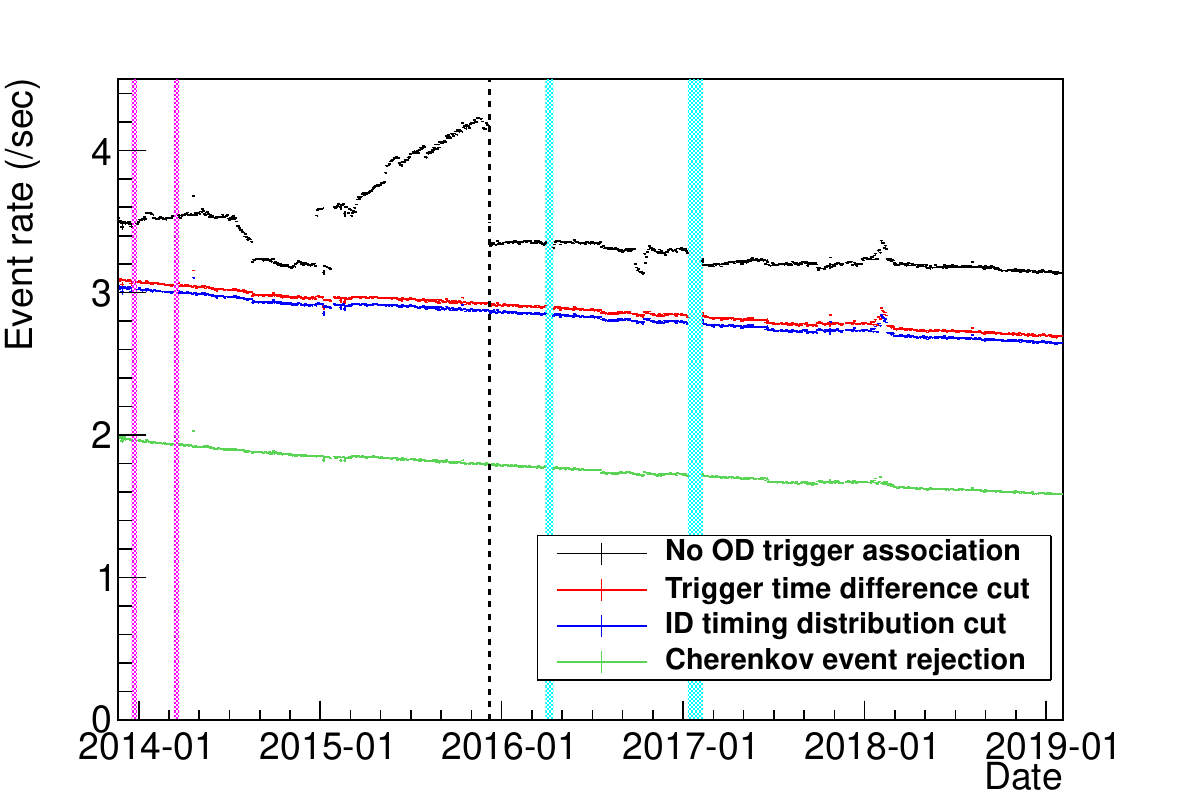}
  \end{center}
  \caption{Time evolution of the normal threshold event rate after the reduction steps of the standard cut.}
  %From top to bottom, the event rate history after the requirement of no OD trigger association (black),
  %the $dT_{\rm pre}$ cut (red), the cut using the ID hit timing distribution (blue),
  %and the Cherenkov event rejection (green) are shown.}
  \label{fig:event-rate}
\end{figure}

Figure~\ref{fig:event-rate} shows time evolution of the XMASS-I normal threshold event rate and the effect from the above selection criteria.
%Low threshold data, which was only used for multi-GeV WIMP searches by modulation analysis, uses additional selection criteria after this standard cut, which will be detailed in Sec.~\ref{Sec:modulation}.

\section{WIMP search in XMASS-I fiducial volume analysis}
\label{Sec:FV}
Given that XMASS-I data was dominated by background originating from $\beta$ events at the ID PMTs entrance window seal at 30~keV$_{ee}$ or less, a fiducial volume (FV) analysis was developed in \cite{XMASS_FV} to reduce this background.
%to address this issue.
% That paper also describes the detailed analysis of all background components contributing events in the fiducial volume, and here we use largely the same background model and evaluation of systematic errors, extended only in time.
% This section describes the results of a WIMP search in a fiducial volume. The background derived from the radioactivities generated in the detector and these systematic errors are estimated in detail. Signals derived from WIMPs are searched by fitting the energy distributions of the data events, BGMC events, and WIMP MC events that remain in the fiducial volume selection. The energy range for exploring WIMPs is 2 to 15 keV. 
% In addition to the standard cuts described before,
%As in that previous analysis a radius $R(T)$ for the event in our almost spherical ID is calculated for each event based solely on hit timings. The nominal position of the event in the ID and with it that position's radius $R(PE)$ on the other hand is based on the event's PE distribution over the ID PMTs; see \cite{XMASS_FV} for details. 
Following~\cite{XMASS_FV}, we used two different position reconstruction methods. One is based on hit timings~\cite{XMASS_RT}, and the other is based on PE distribution~\cite{XMASS_det}. These methods are referred to as $R(T)$, and $R(PE)$, respectively. We used radii $R(T) <$ 38 cm and $R(PE) <$ 20 cm for the FV selection. The fiducial target mass in this volume is 97~kg. The systematic error associated with this reconstruction is discussed in Sec.~\ref{sec:BGsys}.

%As before our volume analysis used events with $R(T) <$~38~cm and $R(PE) <$~20~cm; the fiducial target mass in this volume is 97~kg.

% The dataset and the event selection by standard cut are discussed in section~\ref{sec:Dataset}. In addition, in this analysis, fiducial volume cut by reconstruction based on timing ($R(T)$) and PE ($R(PE)$) observed by PMT are carried out. $R(T)$ and $R(PE)$ was estimated by the comparison of the observed and expected first hit timing or NPE distribution in all PMTs about each position in the detector inside. Both reconstructions are calculated using maximum likelihood method. For $R (PE)$ reconstruction, the optical distribution (reconstruction grid) of the template used for reconstruction is replaced according to the change over time of the optical parameter of LXe. The events, $R(T) <$ 38 cm and $R(PE) <$ 20 cm are required for the event selection, where $R(T)$ and $R(PE)$ are the distance from the detector center to the reconstructed vertex by timing reconstruction \cite{XMASS_takeda} and PE based reconstruction \cite{XMASS_det}, respectively. The fiducial mass of natural xenon in that volume is 97 kg.
% ogawa (220720) add reconstructed energy

\begin{figure}[htbp]
  \begin{center}
	%\vspace{5cm}
    %\includepdf[pages=-, scale=0.5]{paper_spec_reduc.pdf}
    \includegraphics[width=9cm]{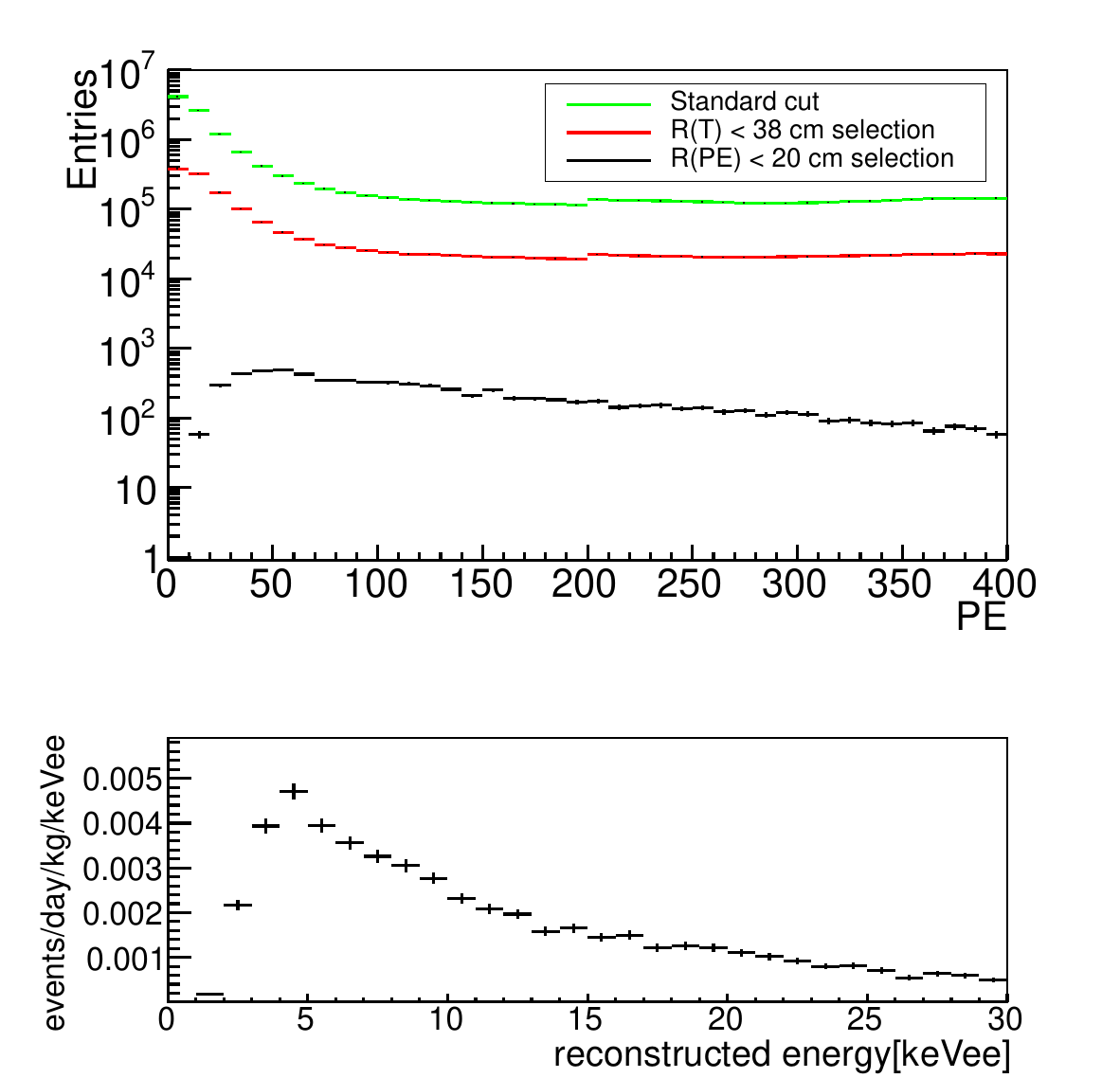}
  \end{center}
  \caption{(Top) PE distributions for each reduction steps after standard cut. %(green) and the two radius cuts $R(T) <$ 38~cm (red) and $R(PE) <$ 20~cm (black). 
  (Bottom) The reconstructed energy spectrum of data after all reductions. Reconstruction details have been explained in~\cite{XMASS_det,XMASS_FV,XMASS_RT}.}
  %\color{red}{NPE is not defined anywhere - just use PE %throughout for number of PE, as defined in the introduction %(photoelectron equivalent [signal].}=>modified (Ogawa,220522)
  \label{fig:FV:reduction}
\end{figure}

Figure~\ref{fig:FV:reduction} shows the selected events' PE distribution before and after these two cuts in the upper panel. The lower panel in Fig.~\ref{fig:FV:reduction} shows the reconstructed energy distribution of the % PE base,
surviving events, taking into account the position dependence of the PE distribution.
A drop in event rate below 4 keV$_{\rm ee}$ reflects the reduced reconstruction efficiency of the PE based reconstruction at lower energies. 
 % for each selection criteria and reconstructed energy distribution after fiducial volume selection. The event rate decreases at 5 keV or less as the energy decreases.  It is introduced by the low efficiency of R(PE) reconstruction. 

\subsection{Background estimation}
\label{sec:BGest}
Here we briefly review the radio-isotopes (RI) considered in our analysis as %contributing
sources of background in the FV; for details see section 4 in~\cite{XMASS_FV}.
Below 30 keV$_{\rm ee}$ the background remaining after the above cuts was dominated by RI at the ID's inner surface facing the LXe target mass. These events are called “mis-identified events" % because the events that occurred in the blind spot of the sensitivity region of PMTs were mistakenly reconstructed in the fiducial volume. 
%as their true origin is obscured by 
as their proper position reconstruction is prevented by the fact that light emitted at the inner OFHC copper surface has no direct path to the nearest PMTs' photocathodes, leading to their PE-based reconstruction being drawn into the FV. 
Candidates for such RI were $^{238}$U, $^{235}$U, $^{232}$Th, $^{40}$K, $^{210}$Pb and their progeny in and on the ID's inner surface materials. 
Detector elements directly facing the LXe at the ID's inner surface were the PMTs' entrance windows, the OFHC copper rings around each ID PMTs' entrance window/metal body connection, and the OFHC copper plates covering these copper rings in the direction of the detector center and in particular also the gaps between the rings of neighboring PMTs.
Below these was the massive OFHC copper structure that supported the PMTs, held them in their position, and further shielded the ID's LXe target volume
from external gammas. 
%aluminum that seals our ID PMTs' quartz window to their metal bodies - this aluminum was found to be contaminated with the upper part of the $^{238}$U decay chain and $^{210}$Pb.

%First, the RI component, which is the background origin of the XMASS detector, will be described. 

%The detail of the BG component in fiducial volume had been already discussed in \cite{XMASS_FV}. 
%Dominant BG component in the analysis energy region $<$30keV is the RIs which originate the inner surface of the detector. The candidate of RI is that $^{238}$U, $^{235}$U, $^{232}$Th, $^{40}$K and $^{210}$Pb in the detector surface material composed of PMTs, copper plate and copper rings for PMT support structure and copper holder for PMTs. These detail structures are explained in \cite{XMASS_FV}. The PMT copper rings are installed to cover the aluminum used to seal the PMT quartz window and metal body, which contains a lot of $^{238}$U upstream and $^{210}$Pb. The copper plates are installed to eliminate the gap made of copper ring. 

All detector materials - except the LXe target material itself - were assayed in high purify germanium (HPGe) detectors~\cite{XMASS_HPGe} and crucial inner surface materials also in a high efficiency surface alpha counter~\cite{XMASS_alpha}.
%The analysis of alpha events in the detector, estimated fram data taken during the first 15 days of data taking, also contributed to our background evaluation. 
Our background model was verified against the first 15 days of data taken and subsequently applied to all data, accounting for the decays of in particular $^{60}$Co ($t_{1/2} = 5.27$ yr) and $^{210}$Pb ($t_{1/2} = 22.2$ yr)~\cite{XMASS_FV}.
% Also the analysis of alpha event in the detector and full volume spectrum can estimate the RI activities. This analysis was performed using the data from the first 15 days of the dataset. The BG MC considers the long-term reduction of events in the 5-year dataset due to the collapse of RI. The significant effect is on the number of events originating from $^{60}$Co ($t_{1/2} = 5.27 yr$) and Pb210 ($t_{1/2} = 22.2 yr$). After the fiducial volume cut, some events generated detector surface are mis-identified as the event in the fiducial volume in the event reconstruction. 

%RIs dissolved in the LXe are asummed. These events are distributed in LXe as uniformly and can not remove using fiducial volume analysis. In this analysis, the impurity in LXe like $^{222}$Rn, $^{85}$Kr, $^{39}$Ar and $^{14}$C are assumed. The $^{222}$Rn and $^{85}$Kr activity was obtained by looking for the coincidence tagging events in the full volume of the ID. The concentration of $^{39}$Ar and $^{14}$C are suggested by the gas analysis of xenon sampling and spectrum fitting result for $>$ 30keV. The condition of the gas circulation changes the RI concentration in LXe. Then the periodically fitting was examined \cite{XMASS_DEC2018}. The other component is the xenon isotope for $^{136}$Xe which make  2$\nu\beta\beta$ spectrum and thermal neutron origin for $^{125}$I, $^{131m}$Xe and $^{133}$Xe are neglisible in this analysis.   
Some RI were found to be dissolved in the LXe itself; their distribution within the target material was assumed to be uniform. 
For this kind of background $^{222}$Rn and its daughter nuclei, $^{85}$Kr, $^{39}$Ar and $^{14}$C were considered in our analysis. 
$^{222}$Rn and $^{85}$Kr concentrations were measured using coincidences in their respective decay chains as observed in the XMASS-I detector itself.
%$^{39}$Ar and $^{14}$C were assayed by analyzing gas samples from the detector and spectral fitting above 30~keV$_{\rm ee}$, respectively. (Ogawa,220913) comment-out because we cannot identify Ar39 by analysing gas sample. also sentence is overlapped with L65. 
An argon contamination found in measurements of xenon gas samples from the detector volume using gas chromatography–mass spectrometry (GC–MS) was subsequently reduced in the distillation campaign.
We also assumed that carbon-containing impurities contaminated the xenon, and the amount of 
$^{39}$Ar and $^{14}$C were determined by fits to the energy spectrum above 30~keV$_{\rm ee}$.

\begin{figure}[htbp]
  \begin{center}
	%\vspace{5cm}
   \includegraphics[width=9cm]{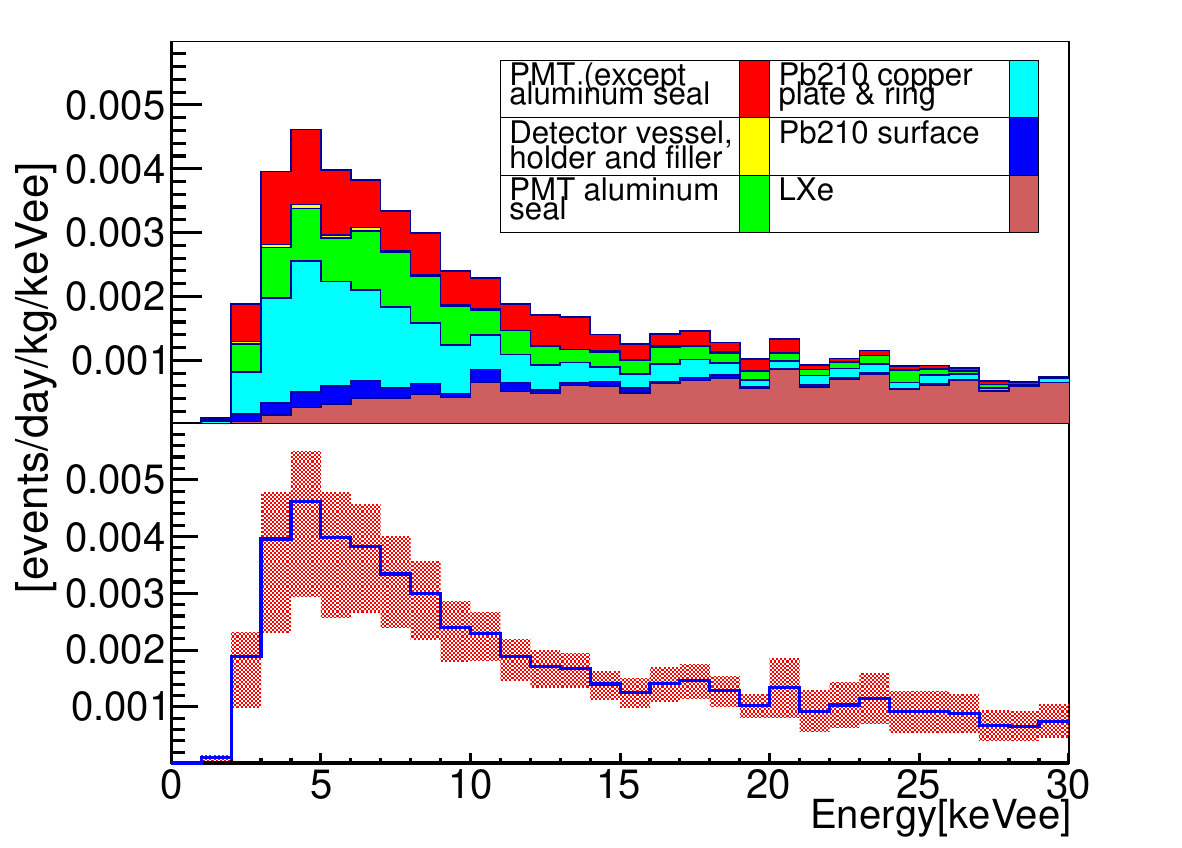}
  \end{center}
  \caption{Energy distributions of the BG simulation after event selection from 0 to 30 keV$_{ee}$ (Top) Colored stacked histograms showing the contributions from various detector components. (Bottom) A cumulative energy spectrum showing the XMASS-I BG-MC with its systematic error band.}
  \label{fig:FV:BGspec}
\end{figure}

% A BG MC simulation are processed with the same livetime in the period of the data set. BG MC reproduces the type and amount of RI evaluated in this section in the detector member. In addition, the change over time in the optical parameter (absorption and scattering length) of liquid xenon was traced, and the same reduction process (standard cut, $R (T)$ cut and $R (PE)$ cut) as the data was performed. The upper figure of Fig~\ref{fig:FV:BGspec} shows the energy distribution of BG for each RI component remaining after fiducial volume selection. Events other than the LXe origin are from the detector surface and are reconstructed within the fiducial volume. These events are called "mis-identified events" because the events that occurred in the blind spot of the sensitivity region of PMT were mistakenly reconstructed in the fiducial volume. 
Our MC simulation of these background sources (BG-MC) used the quantitative evaluations outlined above and traced changes related to changes in the 
Xe circulation pattern over the lifetime of the full data set. 
The simulation also took into account changes of optical parameters 
(the PE yield and the absorption length shown in Fig.~\ref{fig:stability}, and the scattering length)  
as traced in our calibrations. 
 As already mentioned their expected change of rate due to decay was properly taken into account for the isotopes $^{60}$Co and $^{210}$Pb. 
The BG-MC output was then processed using the same selection criteria as was used for the data. 
Figure~\ref{fig:FV:BGspec} shows the spectral composition of our BG-MC for its selected FV events in the upper panel. 
The systematic error of the quadratic sum of all components of this BG-MC evaluation is also shown as the red band in the figure's lower panel, and will be further discussed in Sec.~\ref{sec:BGsys}. 

% \subsection{The correction due to dead PMTs}
%\subsection{Dead PMT correction}
% \subsection{Background estimation due to dead PMT}
\subsection{Dead PMT induced FV events}
\label{sec:deadPMT}
We also describe how we correct for events pulled into the FV due to missing information from dead PMTs. This correction was carefully updated for this analysis over what was used in \cite{XMASS_FV} to address the increase of number of dead PMTs in the later part of the dataset as shown in Fig.~\ref{fig:dead-pmt}.
The dead PMT correction %refers to accumulations of events that form in front of dead PMTs - PMTs that were switched off in the DAQ - in the XMASS-I detector.
originates from the observation that our event reconstruction accumulated excess events after selection in front of dead PMTs, a phenomenon that can be mimicked and verified in data by masking active PMTs in a reconstruction. 
These excess events presumably originate from $\beta$ events near a dead PMT's entrance window seal, but given that the PMT is not read out by the DAQ any more, no signal is recorded on that particular PMT itself.
In the PE reconstruction, which the FV cut was based on, this missing information pulls them into the FV where they appear in front of the dead PMT. 
Since dead PMTs are also taken into account in the XMASS MC, ideally the effect should be reproduced. 
However, there is a difference in the strength of the attracting effect between the results of reconstructions on events with masked PMT information in real data and similar treatment in XMASS MC data. We speculate that this difference is due to inadequate optical modelling of the PMT.
%neighborhood in the XMASS MC. 
%In this section, the ’dead PMT correction’ is described to correct for this difference.
%This phenomenon appears in both data and BG-MC. 
%While XMASS MC was basically tuned by sample data, mainly calibration data, 
%The reproducibility of XMASS MC, except for dead PMT, has been verified by comparison of calibrations, etc., 
%and deviations between data and MC are assessed as systematic errors as discussed in Sec.~\ref{sec:BGsys}, the dead PMT correction for BG-MC simulation was estimated using dead PMT effect factor extracted from observed data. 
%The XMASS MC considers dead PMT but is insufficient the degree to pull into the FV in the reconstruction. 

A correction factor for the BG-MC spectrum is applied for such differences in each of the energy regions 2-5, 5-15, 15-20 and 20-30~keV. These correction factors were estimated by comparing of the distance between the projection of the reconstructed vertex onto the detector surface and the dead PMT position ($D_{dead}$) between data and BG-MC in the fiducial volume. 
In smaller $D_{dead}$ region the effect due to dead PMT increases and in larger $D_{dead}$ region the effect decreases.
In the correction calculation, to estimate effects from other than dead PMT, the $D_{dead}$ region where the effect of dead PMT becomes negligible was estimated.
The number of events of the data and MC in this region was used to normalize the $D_{dead}$ distributions and then differences in distributions were evaluated. 
This region for normalization was estimated as $D_{dead}>$20~cm. 
Since the $D_{dead}$ distribution naturally depends on the number of dead PMTs, it was required to update this boundary value from the previous analysis.

Three different systematic errors were associated with this dead PMT correction. The first stemmed from the statistical uncertainty of the correction factor itself.
%The second came from the difference between the correction factor, thus establishing an expected correction from an actual increase in reconstructed FV events when well working PMTs were masked in our analysis to study this effect.
The second contribution was estimated by the difference in the correction factor estimated from the systematic difference of event rates in the fiducial volume by deliberately masking normal PMTs.
The third systematically considered the possibility that the dead PMT effect reached farther than the 20 cm boundary used in our estimation.

\begin{figure}[htbp]
  \begin{center}
	%\vspace{5cm}
  \includegraphics[width=9cm]{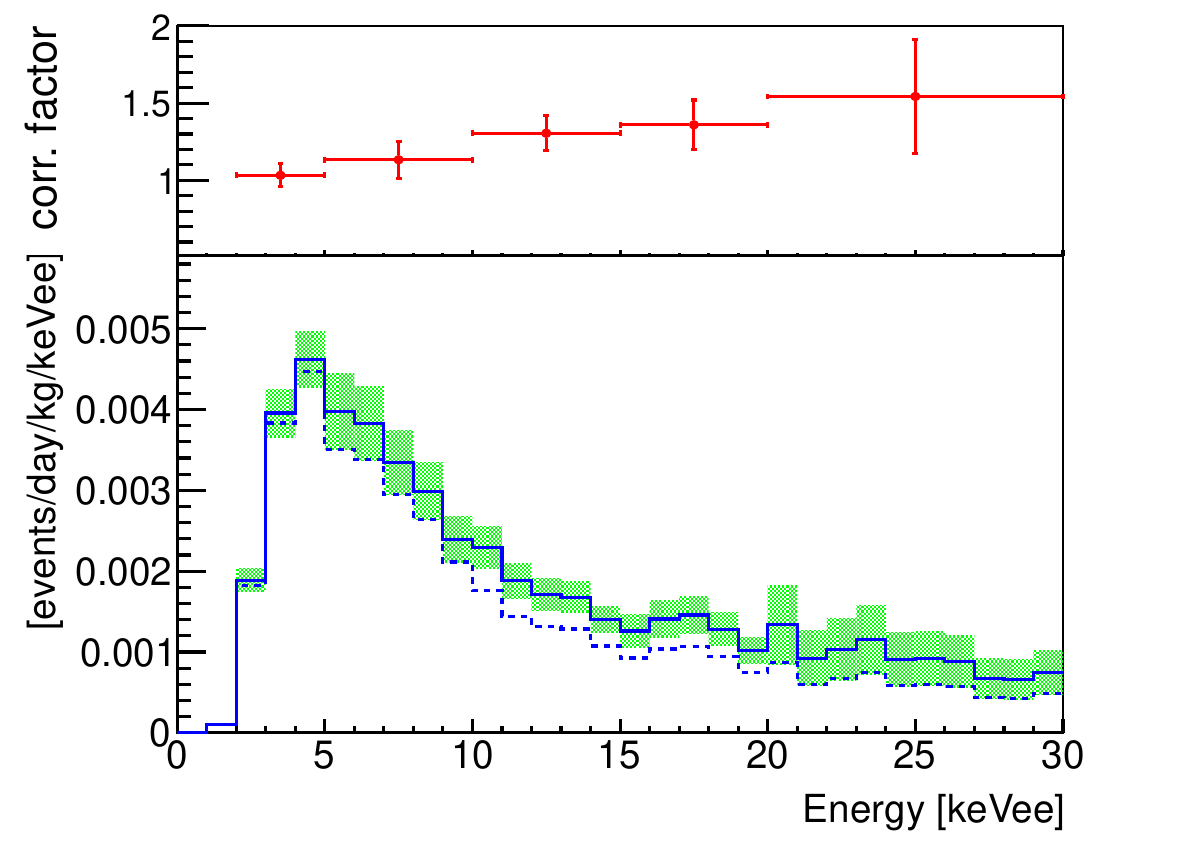}
  \end{center}
  \caption{(Top) The dead PMT correction factor with its associated systematic error. (Bottom) The total BG-MC energy spectrum before (dotted) and after (solid) applying the dead PMT correction. The green band shows the range of this correction's systematic uncertainty.}
  \label{fig:FV:deadtube}
\end{figure}

%These three systematic errors affect the validity for the value of the dead PMT correction. 
Figure~\ref{fig:FV:deadtube} shows the actual correction factor with its the associated systematic error at the upper panel. The lower panel shows the total BG-MC after this correction is applying and its associated total systematic uncertainty reflected by the underlying green band. %of dead PMT correction overlaid as a green band.
%Table~\ref{tab:FV:sysBG} below details the various contributions to the resulting overall systematic uncertainty of our BG-MC.
% \begin{figure}[htbp]
%  \begin{center}
%	%\vspace{5cm}
%   \includegraphics[width=9cm]{distance_paper.pdf}
%  \end{center}
%  \caption{The distance distribution between the extension of reconstructed vertex to detector surface and dead tube position for data and BG MC for 5-30keV. Black histgram is data and red histogram is BG MC. The event both of data and BG MC apply the standard cut, $R(T) <$ 38cm selection and $R(PE) <$ 20 cm selection. Green line is the uniform distributed toy MC.}
 % \label{fig:FV:distance}
%\end{figure}

\subsection{Systematic errors in BG-MC}
%title of subsection was changed 220723 Ogawa
\label{sec:BGsys}
% The evaluation of systematic errors for our BG-MC is basically same with WIMPs search analysis ~\cite{XMASS_FV}, except for the one related to the dead PMT correction. 
%PMT contains a large amount of RI ($^{210}$Pb and the upper stream of $^{238}$U) in aluminum for sealing the quartz window and metal body ~\cite{XMASS_det}. The copper rings are attached to the PMT window/metal body transition to reduce the signal from these RIs. In addition, the copper plates are installed on the ID inner surface to eliminate the gap caused by the copper ring. The events that remain in the fiducial volume selection are subject to the indeterminacy associated with these structures.
% Table~\ref{tab:FV:sysBG} shows the changing in BGMC simulation given by the each contents of the systematic uncertainty. The contents of (1) to (5) are due to the structure of the detector surface. They use XMASS MC to find the indefiniteness of the fiducial volume selection that accompanies the indefiniteness of the structure. (6) is the indefiniteness of reconstruction itself. (7) is an indeterminacy related to the emission time constant of scintillation light. (8) is the indefiniteness of the dead PMT origin. (9) is the indeterminacy due to the effect of liquid xenon on the reconstruction of the optical property discussed in ~\ref{sec:Calib}.

% Except for the dead PMT correction the evaluation of the systematic errors for this analysis proceeded as described in our previous publication~\cite{XMASS_FV}. The systematic error evaluation for the dead PMT correction was discussed above in~\ref{deadPMT}.

The lower panel of Fig.~\ref{fig:FV:BGspec} shows the BG-MC spectrum with its total systematic error as the quadratic sum of all its components. % of BG-MC. 
Table~\ref{tab:sysBG} summarizes this analysis' systematic error estimates for the energy ranges relevant to this analysis. Its rows are organized thematically: 
%(1) through (5) relate to the OFHC copper structures installed at the inner surface of the ID after the XMASS commissioning phase to deal with the RI contamination in the PMTs' aluminum seal between its metal body and quartz entrance window as detailed in ~\cite{XMASS_FV}, (6) through (9) to our reconstruction and analysis methods. 
% the contents of 
(1) to (5) were due to the structure and properties of the ID's inner surface. Therefore, the respective geometries were varied in the MC to explore the relevant range and obtain these values.
(6) quantified effects of our PE-based position reconstruction, in particular that of the FV cut, and 
(7) the uncertainty in the scintillation light's emission time constant. %of scintillation light. 
(8) stemmed from the dead PMT correction as explained above in Sec.~\ref{sec:deadPMT}. 
(9) summarized the uncertainty in the LXe optical properties as reviewed in Sec.~\ref{Sec:Calib}.
These systematic errors are not symmetric, because the number of events in the FV does not increase or decrease symmetrically with a change in these parameters as reconstructed event positions depend on them. 
%The events in FV do not increase or decrease symmetrically, but asymmetrically. It is because it assesses how the events in the FV increase or decrease by changing the state parameter of the detector, which is independently assessed.

\begin{table*}[t]
% \caption{The changing in BG MC simulation given by the each contents of the systematic uncertainty.}
\caption{BG-MC simulation systematic error by each content. Ring means a copper ring around each PMTs’ window/metal. See Fig.~1(b) in ~\cite{XMASS_FV} for detail. As the surface roughness of this ring affects the optical reflection property, a systematic error was introduced to account for its uncertainty. The timing response includes both the scintillation decay time and the fluctuation in the PMT.}
\label{tab:sysBG} 
\begin{center}
\begin{tabular}{lcccc}
    \hline \hline
%  \multicolumn{1}{l}{The contents of uncertainty} & \multicolumn{4}{c}{The value of uncertainty} \\
  \multicolumn{1}{l}{Contents} & \multicolumn{4}{c}{Evaluated systematic errors} \\
	& 2-5 ${\rm keV_{ee}}$ & 5-10 ${\rm keV_{ee}}$ & 10-15 ${\rm keV_{ee}}$ & 15-30 ${\rm keV_{ee}}$\\
    \hline
%(1) Plate gap & +9.1/-33.4$\%$ & +5.2/-19.1$\%$ & +3.1/-11.3$\%$ & +1.6/-6.0$\%$\\
(1) Gaps between adjacent plates & +9.1/-33.4$\%$ & +5.2/-19.1$\%$ & +3.1/-11.3$\%$ & +1.6/-6.0$\%$\\
(2) Ring roughness & +9.7/-10.3$\%$ & +5.6/-5.9$\%$& +3.3/-3.5$\%$&+1.8/-1.9$\%$\\
(3) Cu reflectivity & +3.6/-0.0$\%$ & +5.9/-0.0$\%$& +4.4/-0.0$\%$&+2.4/-0.0$\%$ \\
%(4) Plate floating & +0.0/-6.7$\%$ & +0.0/-3.8$\%$ &+0.0/-2.3$\%$&+0.0/-1.2$\%$\\
(4) Unevenness due to thin plate buckling & +0.0/-6.7$\%$ & +0.0/-3.8$\%$ &+0.0/-2.3$\%$&+0.0/-1.2$\%$\\
(5) PMT aluminum seal& +1.0/-1.0$\%$ & +0.3/-0.3$\%$&+0.0/-0.0$\%$& +0.0/-0.0$\%$\\
(6) Reconstruction & +8.9/-8.9$\%$ & +1.4/-7.8$\%$ & +2.8/-2.8$\%$ &+2.8/-2.8$\%$\\
(7) Timing response& +3.1/-9.9$\%$ & +7.6/-11.3$\%$ &+0.4/-5.3$\%$&+0.4/-5.3$\%$\\
(8) Dead PMT & +7.5/-7.5$\%$ &  +11.9/-11.9$\%$ &+11.4/-11.4$\%$&+28.3/-28.3$\%$\\
(9) LXe optical property & +0.9/-6.7$\%$ & +0.9/-6.7$\%$ &+0.8/-6.7$\%$&+1.5/-1.1$\%$\\
    \hline \hline
\end{tabular}
\end{center}
\end{table*}

\subsection{Results and discussion}
\label{sec:FV:results}
% The WIMP signal was searched by fitting the observed energy spectrum with the sum of evaluated BG and the signal. 
To search for a potential WIMP signal the observed energy spectrum was fit as the sum of the corrected BG-MC and a simulated signal contribution of unknown size.
For the WIMP signal, WIMP-nucleus elastic scattering events were simulated for WIMP masses from ${\rm 20\,GeV/c^{2}}$ to ${\rm 10\,TeV/c^{2}}$. For these simulations we assumed the parameters usually used to report such results: the standard spherical and isothermal galactic halo model with a solar system speed of ${\rm v_{0}=220\,km/s}$, a Milky Way's escape velocity of ${\rm v_{esc}=544\,km/s}$ \cite{LewinSmith2}. and a local DM halo density of ${\rm 0.3\,GeV/cm^{3}}$, following Ref.~\cite{LewinSmith}. 
The same event selection that was applied to data and BG-MC was also applied to the WIMP MC.

In the fits of the data to the sum of the dead PMT corrected BG-MC plus the simulated WIMP response of the detector for a specific WIMP mass in the energy range of 2--15~keV${_{ee}}$ for WIMPs we used the following $\chi^{2}$ definition:

\begin{equation}
\chi^{2} = \sum_{i}\frac{ ( D_{i} - B_{i}^{*} - \alpha \cdot W_{i}^{*} )^{2} }{ D_{i} + \sigma^{*}(B_{stat})_{i}^{2} + \alpha^{2} \cdot \sigma(W_{stat})_{i}^{2}} + \chi_{pull}^{2}\\.       
\end{equation}
\begin{equation}
B_{i}^{*} = \sum_{j} p_{j}( B_{ij} + \Sigma_{k} q_{k} \cdot \sigma(B_{sys})_{ijk} ), \\
\end{equation}
\begin{equation}
W_{i}^{*} = W_{i} + \sum_{l} r_{l} \cdot \sigma(W_{sys})_{il}, \\
\end{equation}
\begin{equation}
\sigma^{*}(B_{stat})_{i}^{2} = \sum_{j} p_{j}^2 \cdot \sigma(B_{stat})_{ij}^{2}, \\
\end{equation}
\begin{equation}
\chi_{pull}^{2} = \sum_{j}\frac{(1-p_{j})^2}{\sigma(B_{RI})_{j}^{2}} + \sum_{k}q_{k}^{2} + \sum_{l}r_{l}^{2},
\end{equation}
where $D_{i}$, $B_{ij}$, and $W_{i}$ are the number of events in the data, the BG estimate, and WIMP MC simulations, respectively. $i$ enumerated the energy bins from 2 to 15~keV${_{ee}}$ (13 bins).
$\alpha$ is free parameter in the fit and a scaling factor for the WIMP MC contribution corresponding to the WIMP-nucleon cross section. Therefore, the number of degrees of freedom (n.d.f.) for the fit was 12.  
The variable $j$ enumerated the BG sources (20 components) in the BG-MC~\cite{XMASS_FV}.
The variables $k$ and $l$ enumerated the different systematic errors in the BG estimate (9 components) and WIMP MC simulations (4 components), respectively.
${\sigma(B_{stat})_{ij}}$ and ${\sigma(W_{stat})_{i}}$ are the statistical uncertainties in the BG estimate and the WIMP MC simulations, respectively.
$\sigma(B_{RI})_{j}$, $\sigma(B_{sys})_{ijk}$, and $\sigma(W_{sys})_{i}$ are uncertainties in the amount of RI activities, systematic errors in the BG estimate (Table~\ref{tab:sysBG}) and the WIMP MC simulations, respectively.
 $p_{j}$, $q_{k}$, and $r_{l}$ are scale factors for the amount of RI activity, the systematic errors in the BG estimate, and the systematic errors in the WIMP MC simulations, respectively. They were varied with constraints from the pull term shown in the equation (5) in finding minimum chi-squared value. As shown in Table 1, the amount of the systematic errors in the BG estimate ($\sigma(B_{sys})_{ijk}$) is different between positive and negative in most cases. A positive value of $\sigma(B_{sys})_{ijk}$ is chosen if $q_{k} \geq 0$, on the other hand a negative value is chosen if $q_{k} < 0$.

\begin{figure}[htbp]
  \begin{center}
	%\vspace{5cm}
    \includegraphics[keepaspectratio=true,height=58mm]{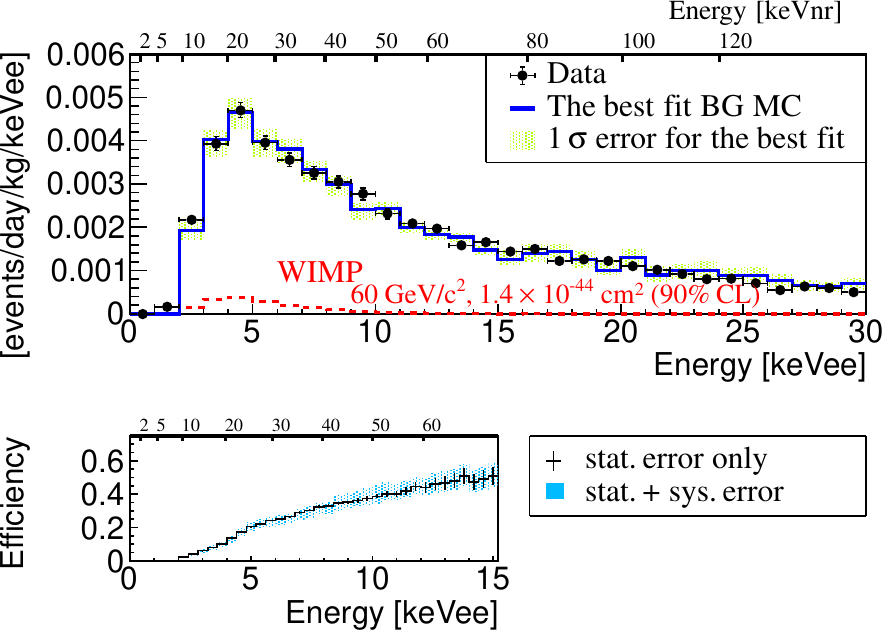}
  \end{center}
  \caption{Data spectra with the statistical error shown in filled dots, and the BG estimate shown in a thick line with the $1\sigma$ error from the best fit shown as a shaded band with an energy region between ${\rm 2\,keV_{ee}}$ and ${\rm 30\,keV_{ee}}$ (top). The WIMP MC expectation for ${\rm 60\,GeV/c^{2}}$ is also shown in a dotted line with energy region between ${\rm 2\,keV_{ee}}$ and ${\rm 15\,keV_{ee}}$. Energy in keVnr is shown on the top. The bottom shows the signal efficiency after applying same event selections.}
  \label{fig:FV:FittingResults}
\end{figure}

Signal efficiency is defined as the number of retained WIMP events after applying event selections divided by the number of WIMP events generated in the FV of the detector. 
Systematic uncertainties for signal spectrum prediction were evaluated as shown in Ref.~\cite{XMASS_FV}.
The largest systematic error came from the uncertainty in the scintillation decay time of ${\rm 26.9^{+0.8}_{-1.2}\,ns}$ for NR, which had been updated to reflect the latest results from our ${\rm ^{252}Cf}$ neutron calibrations \cite{XMASS_neutron}. Since the Cherenkov cut affects some of the expected WIMP signal, %was affected by our Cherenkov event rejection in the standard cut, signal efficiency was affected by the change in the decay time this update brought about.
the change in this decay time also changed the signal efficiency. 
Efficiencies of the WIMP signal after applying the standard, $R(T)$, and $R(PE)$ cuts, %were calculated: for ${\rm 60\,GeV/c{2}}$ WIMPs they were $12\%$, $32\%$, and $48\%$, averaged over the energy ranges 2-5, 5-10, and 10-15${\rm\,keV_{ee}}$, respectively; the fit result for ${\rm 60\,GeV/c{2}}$ WIMPS is shown in Fig.~\ref{fig:FV:FittingResults}.
averaged over the energy ranges 2--5, 5--10, and 10--15${\rm\,keV_{ee}}$ were $12\,\%$, $32\,\%$, and $48\,\%$, respectively, for ${\rm 60\,GeV/c^{2}}$ WIMPs. The spectrum fit for these ${\rm 60\,GeV/c^{2}}$ WIMPs is shown in Fig.~\ref{fig:FV:FittingResults}.
%The energy spectrum of the data are the filled dots and the solid blue histogram reflects the best fit. 
This best fit to our BG estimate plus signal model had a $\chi^{2}$ of 12.4 (n.d.f.=12) with a null WIMP contribution ($\alpha$=0).
As the observed event distribution is thus consistent with our BG evaluation, a $90\%$ confidence level (CL) upper limit on the WIMP-nucleon cross section is calculated such that the integral of the probability density function $\exp(-\Delta\chi^{2}/2)$, where $\Delta\chi^{2}=\chi^2-\chi^2_{min}$, becomes $90\%$ of the total at the limiting cross-section. The red dotted line in this figure corresponds to the signal contribution at that $90\%$ CL upper limit for ${\rm 60\,GeV/c^{2}}$ WIMPs. % is also shown as the red dotted line in Fig.~\ref{fig:FV:FittingResults}. 

% The data energy spectrum was fitted with the sum of the BG estimate shown in Fig.~\ref{fig:FV:deadtube} and the WIMP contribution in the energy range of 2-15${\,keV_{ee}}$ using the following $\chi^{2}$ definition:

% The best fit had a $\chi^{2}$ of 12.4 (n.d.f=12) with a null WIMP contribution ($\alpha$=0). % move it to where it belongs: with the other text regarding this fig
%Figure~\ref{fig:FV:FittingResults} shows the energy spectrum of the data as filled dots and the BG estimates with the solid blue histogram reflecting the best fit. This best fit had a $\chi^{2}$ of 12.4 (n.d.f=12) with a null WIMP contribution ($\alpha$=0).
%All the remaining events are consistent with our BG evaluation, and thus a $90\%$ confidence level (CL) upper limit on the WIMP-nucleon cross section was calculated for so that the integral of the probability density function $\exp(-\Delta\chi^{2}/2)$ becomes $90\%$ of the total at the limiting cross-section. The red dotted line in this figure corresponds to the signal at the $90\%$ CL upper limit for ${\rm 60\,GeV/c{2}}$ WIMPs. % is also shown as the red dotted line in Fig.~\ref{fig:FV:FittingResults}. 

\begin{figure}[htbp]
  \begin{center}
	%\vspace{5cm}
    \includegraphics[keepaspectratio=true,height=90mm]{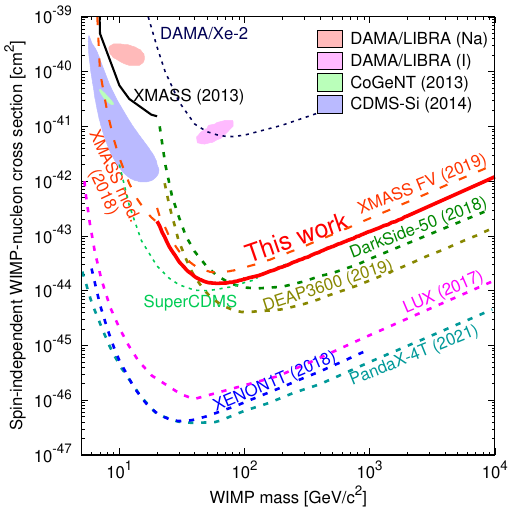}
  \end{center}
  \caption{The spin-independent WIMP-nucleon cross section limit as a function of the WIMP mass at the $90\%$ CL for this work is shown as a solid thick line. Limits ~\cite{XMASS_FV, XMASS_LowMassWIMP, XMASS_Modulation2018, DAMAXe-2, DARKSide, SuperCDMS, DEAP3600, LUX, XENON1T_1ty, PandaX-4T} as well as allowed regions ~\cite{DAMA_LIBRA, CoGeNT_2013, CDMS_Si} from other experimental results are also shown.}
  \label{fig:FV:90CLlimits}
\end{figure}

Such fits of the respective simulated detector response to a WIMP interaction plus the dead PMT corrected BG-MC were done for all simulated WIMP masses, and
the resulting $90\%$ CL upper limits for different WIMP masses are plotted in Fig.~\ref{fig:FV:90CLlimits}. Our lowest limit is ${\rm 1.4\times 10^{-44}\, cm^{2}}$, attained at a WIMP mass of ${\rm 60\,GeV/c^{2}}$ from the fit shown in Fig.~\ref{fig:FV:FittingResults}. This is the most stringent limit among results from single-phase LXe detectors.

% \section{WIMP search by annual modulation}
\section{Annual modulation WIMP search}
\label{Sec:modulation}

The velocity of the earth relative to the galactic DM halo changes as the Earth moves around the sun.
This is because the Earth's velocity in its orbit around the Sun effectively adds to or subtracts from the Sun's velocity through a stationary halo. 
%The Earth's velocity relative to the galactic DM halo changes as the Earth moves around the Sun. The velocity of the Earth in its orbit around the Sun effectively adds to or subtracts from the Sun's velocity through a stationary halo. 
This causes a corresponding change in the expected DM signal rate of terrestrial detectors, with this relative velocity modulation
%of the earthbound detector's target material
affecting signal rates~\cite{Drukier}. 
% In the past we reported searches for dark matter with the XMASS detector by looking for this modulation signal.
Searches using this tell-tale signal rate modulation were already conducted with parts of the XMASS-I data  
% Those are the searches for dark matter mass 
for NRs from multi-GeV 
%WIMP masses (4-20 GeV) for nuclear recoils 
WIMPs~\cite{XMASS_Modulation,XMASS_Modulation2018}, as well as for the sub-GeV mass region % (0.32-1 GeV),
where bremsstrahlung is expected to boost the signal~\cite{XMASS_Modulation2019}.
% In this work, we conducted the search for annual DM signal modulation analysis by using all the data obtained during a long-term measurement.
Here we updated these results using our final five-year data set and full three years of low threshold data, as shown in Table~\ref{tab:data_set}.
A new search exploiting the Migdal effect \cite{Migdal} in the sub-GeV WIMP mass region was added to this paper.

\begin{table}[h]
    \centering
        \caption{Annual modulation %analysis
        data sets and their thresholds}
    \begin{tabular}{c|c|c|c|c}
    \hline
    threshold     &  date & PE& keV$_{\rm{ee}}$& keV$_{\rm{nr}}$  \\
    \hline
    low  &  Dec.8.2015 - Feb.1.2019& 2.3 & 0.5 & 2.3\\
    normal& Nov.20.2013 - Feb.1.2019&  6.0 &  1.0 & 4.8\\
    \hline
    \end{tabular}

    \label{tab:data_set}
\end{table}

For %the purpose of 
these analyses, data were binned in live time intervals of roughly 15 live days per bin, resulting in 125 bins for the normal and 67 bins for the low threshold data.
%For the NR analysis
Both normal and low-threshold triggered events ranging in energy from 0.5 to 20~keV$_{\rm{ee}}$ (2.3 to 99.6 keV$_{\rm{nr}}$) were used for the NR analysis. 
The energy range for both the bremsstrahlung and Migdal analyses %the energy range 
was 1 to 20~keV$_{\rm{ee}}$, using only normal threshold data. 
%This energy threshold of 1 keV$_{\rm{ee}}$ 
The 1 keV$_{\rm{ee}}$ energy threshold for these ER signals 
was set as the uncertainty in the scintillation efficiency for electrons and gamma rays %estimated by the calibration 
increases considerably below that energy.
Though the response below 1 keV$_{\rm{ee}}$ was implemented in the XMASS MC,
the ER signal below 1 keV$_{\rm{ee}}$ was not considered.
The scintillation efficiency and the uncertainty above 1keV$_{\rm{ee}}$ is shown in Fig.~\ref{fig:scint_eff}.  
%For more detail, see~\cite{XMASS_Modulation2019}.

All modulation analyses including NR analysis were done with 
keV$_{\rm{ee}}$ unit.

\subsection{Analysis and results of multi-GeV WIMPs}
The spin-independent NR signal in the energy range from 0.5 to 20 keV$_{\rm{ee}}$ 
(2.3 to 99.6 keV$_{\rm{nr}}$) 
was used to study annual modulation induced by WIMPs in the multi-GeV mass range. 
% Data on energy less then 1.0 keV$_{\rm{ee}}$ were analysed using nuclear recoils
Events at an energy threshold of $\sim$0.5 keV$_{\rm{ee}}$ average a recorded detector response of 2.3 PE. 

%\subsubsection{Data selection}

%Before retrieving time variation information from the data, event reduction was performed  to %reduce background mainly from Cherenkov light in PMT windows and from events near the detector %wall as described in \cite{XMASS_MOD} by applying standard cuts. 
%The count rate for data after all cuts including our likelihood cut is $\sim$0.75 %events/day/kg/keV$_{\rm ee}$ at 1.0  keV$_{\rm ee}$.  
%The signal efficiency was evaluated from MC simulation with events uniformly distributed %throughout the sensitive volume.  In order to estimate the efficiency, a flat energy spectrum was %assumed and the fraction of remaining events after all cuts was calculated.  The bottom panel of  %Fig.~\ref{fig:hist} shows the signal efficiency after all cuts with 832 kg LXe target. Overall our %improved event selection $-$ while keeping the signal efficiency$-$ brings about a further %reduction in data size by about 30\% at low energy compared to our previous publication.

\subsubsection{Additional event and run selections}
\label{Sec:additional}

As explained in \cite{XMASS_Modulation2018},  
an effective background reduction near the ID wall, which was the major background in this analysis, can be achieved by constructing a likelihood function ($L$) based on the sphericity and aplanarity of PE hit patterns, as well as 
the fraction of the PE counts on the ID PMT with the largest PE signal to the total PE count recorded on all ID PMTs in the event.
With $L_{s}$ denoting the likelihood for a signal event uniformly distributed in the target mass and $L_{b}$ the likelihood for a background event near the wall ID's inner surface (wall event), the cut parameter in –ln($L_{s}/L_{b}$) was chosen so that it kept 50\% efficiency for signal after the standard cuts. 
The actual cut parameter, therefore, depends on the total PE in the event. 

\begin{figure}[htbp]
\centering
\includegraphics[width=0.5\textwidth]{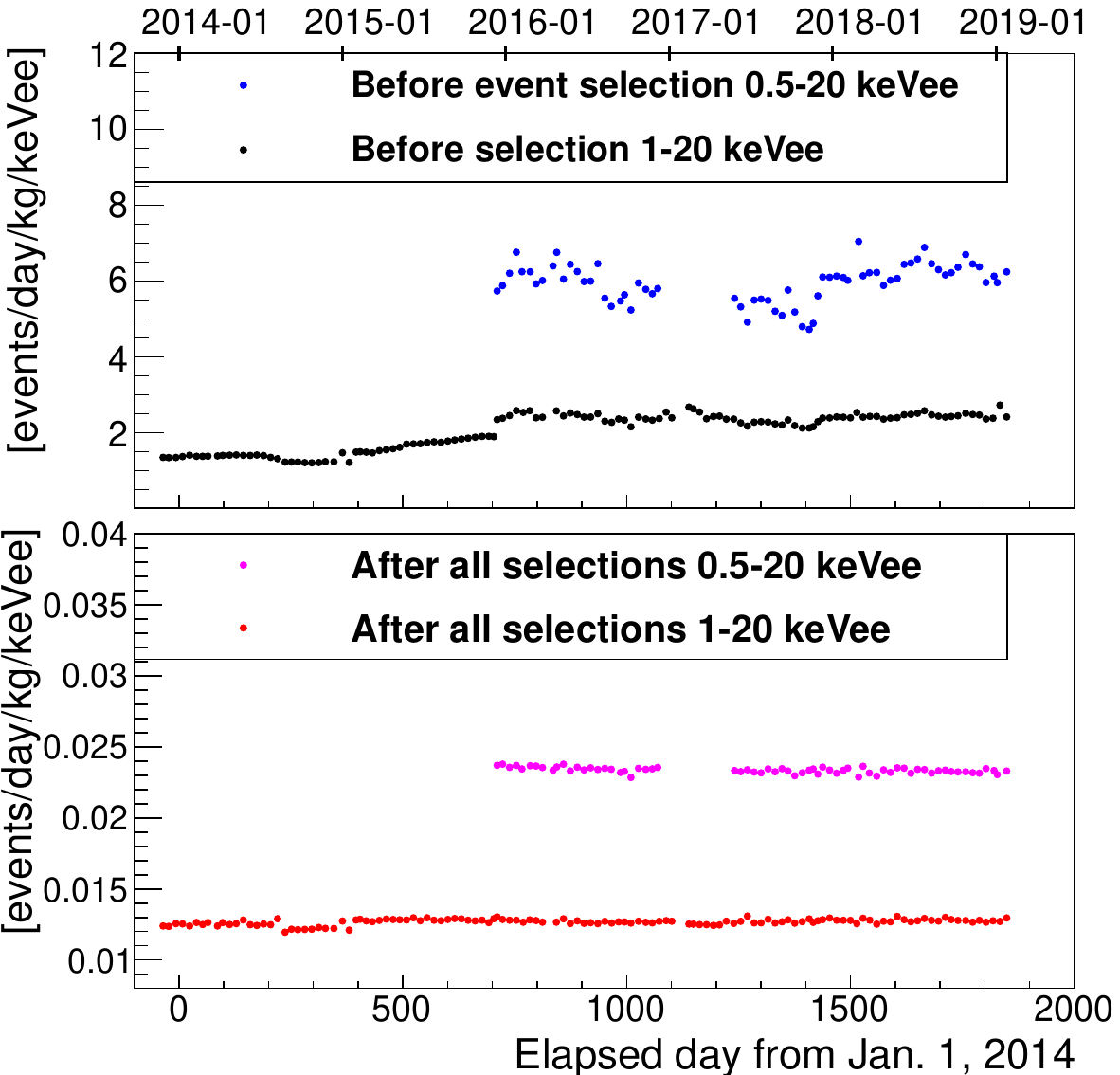}
\caption{Time evolution of the event rate before selection (top panel) and after all selections (bottom panel). Blue and magenta show 0.5--20keV$_{\rm ee}$ event rates and 1--20keV$_{\rm ee}$ is shown by black and red.}
\label{fig:eratelow}
\end{figure}

Another significant background component in the low-threshold data stemmed from the light emission of ID PMTs~\cite{mkobaD}.
Such light emission could be triggered by even only a single PE being released from the photocathode in the emitting PMT.
%recorded on the emitting PMT.
% From the measurement for several PMTs at room temperature,
At room temperature we measured the probability for such emission from a single PE on several PMTs and found it to be in the range of
% the probability of the emission per single PE was 
$\sim$0.3--1.0\%.
Given that this light emission could also occur after dark counts initiated by the thermal emission of an electron from the photocathode, the dark rate of ID PMTs directly affects the event rate at the analysis threshold.
To address this background, 
information from the LED calibration as discussed in Sec.~\ref{Sec:Calib}, 
PMT dark rates and also PMT gains were used for the additional run selection.
Averages and dispersion of these parameters were evaluated for two days periods 
and two day periods with
statistically significant deviations were removed from low-threshold data set.
%in this additional run selection. 
The longest period to be removed was the one after the Xe purification work at the beginning of 2017, 
together with the period rejected by the run selection mentioned in Sec.~\ref{Sec:Data:run}. These removed period can be seen in Fig.~\ref{fig:eratelow} as gaps of 0.5--20keV$_{\rm ee}$ event rate.

% Furthermore, e
Events with this light emission are also characterized by specific relative timing and positioning of the respective hits;
% angular distributions of  PMTs' hits; 
the time difference between the hit caused by such light emission and the hit causing such light emission was larger than 35 ns, and 
the angle between 
%the two lines connecting the positions of these two PMTs and the center of the ID 
the line connecting the positions of these two PMTs and the line connecting the first hit PMT and the center of the ID 
was smaller than 50 degrees.
% former hits. Angle is defined by three points, two hit PMT's photo-cathode centers and the detector center.
%{\bf we need a definition of angle.}
Only for three hit events, an additional condition was placed, requiring that no pair of the three hits met the above conditions on relative timing and angle:
if any pair met these conditions, the event was not used in the analysis.
The contribution from this light emission was negligible for events with 
larger than four hits. 
This additional event selection is referred to as the flasher cut. 

\begin{figure}[htbp]
\centering
\includegraphics[width=0.53\textwidth]{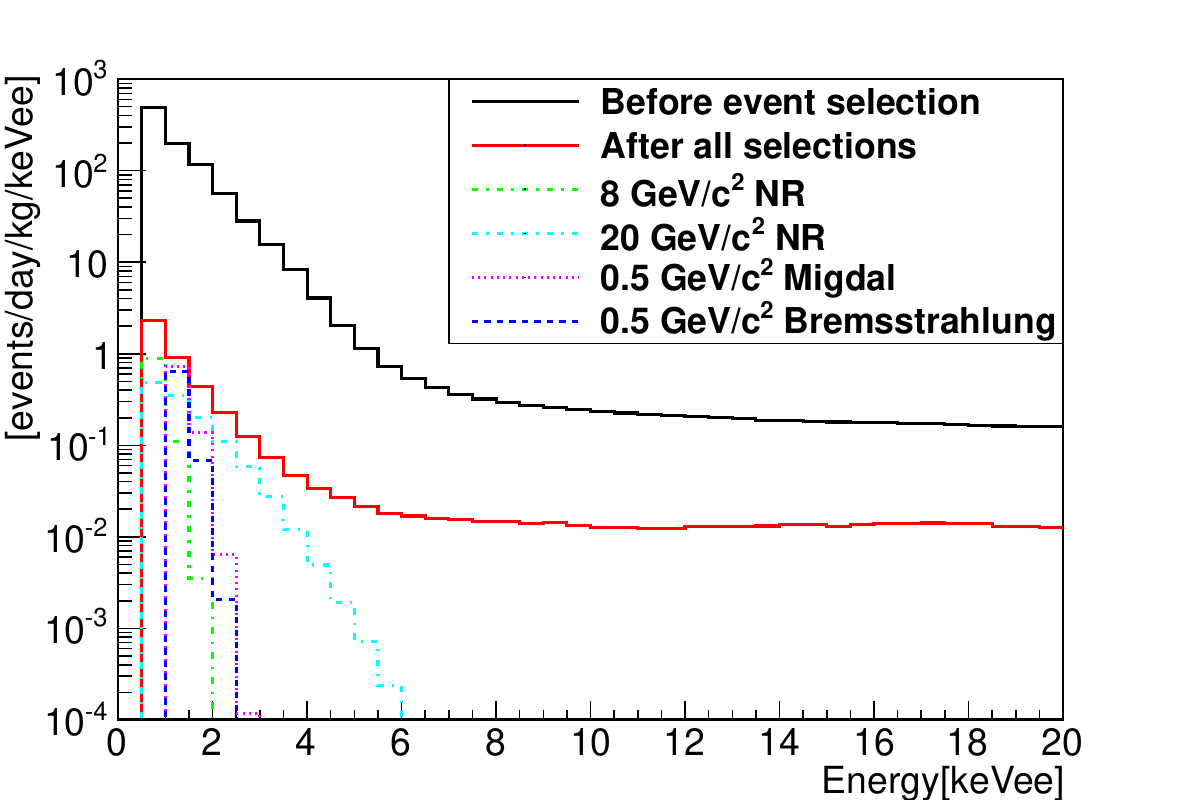}
\caption{Energy spectra of observed data before and after event selection, and those of some simulated signal shapes: NR from 8 and 20~GeV WIMPs with $10^{-41}$, $10^{-42}\rm{cm}^{2}$ cross sections, and a Migdal and a bremsstrahlung simulation for 0.5~GeV WIMPs with cross sections of $6\times10^{-34}\rm{cm}^{2}$ and $3\times10^{-32}\rm{cm}^{2}$, respectively.}
% Simulated signal for Migdal effect (0.5~GeV with $6\times10^{-34}\rm{cm}^{2}$) and bremsstrahlung (0.5~GeV with $3\times10^{-32}\rm{cm}^{2}$) are also included.  
%The figure in top panel 
%also includes  the spectra of some simulated signals from Migdal and Bremsstrahlung effect.}
\label{fig:hist}
\end{figure}

Figure~\ref{fig:hist} shows the energy spectra for the whole data set after all selection criteria have been applied and the spectra of some simulated NR signal spectra for comparison.
The time evolution of the event rate before selection and after all selections are shown in Fig.~\ref{fig:eratelow}.
%at the top
%(the figure also includes the spectra of some simulated signals from Migdal and Bremsstrahlung effect), while the bottom panel shows the overall efficiency for the NR signal.
%The event rate evolution versus time before and after all selections is shown in Fig.~\ref{fig:eratelow}, for 0.5--20~keV$_{\rm ee}$ and 1--20~keV$_{\rm ee}$.

The expected annual modulation amplitude in the DM NR signature was
discussed as in \cite{LewinSmith} and
evaluated in the same way as in our previous XMASS analysis~\cite{XMASS_Modulation,XMASS_Modulation2018,XMASS_Modulation2019}.

\begin{figure*}[htbp]
\centering
\includegraphics[width=1.0\textwidth]{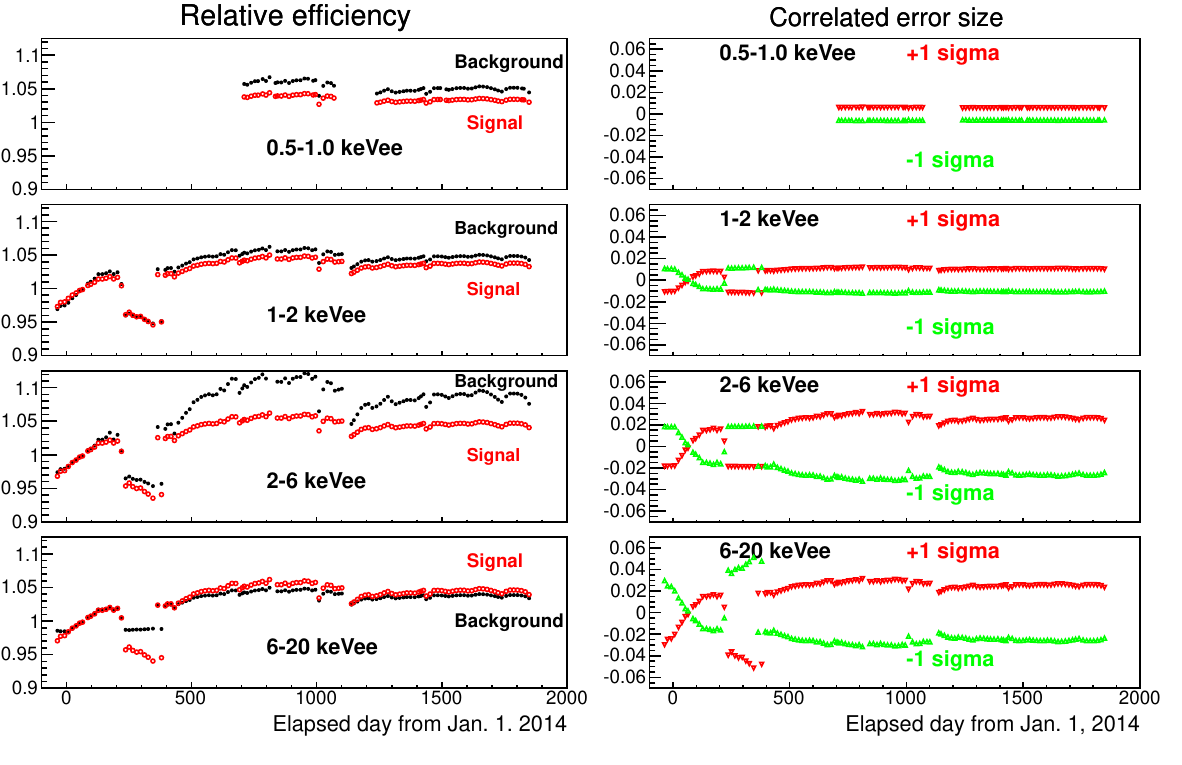}
\caption{ 
(Left) Relative efficiency mean values for both signal (red open circle) and background (black closed circle) events: the overall efficiency was normalized at an absorption length of 8 m for all energy ranges.\\
(Right) 1$\sigma$ ranges of uncertainty in the BG relative efficiencies shown in the left panel. The zero crossing near day 80 is where the improving absorption length passed 8~m.
}
\label{fig:sys}
\end{figure*}

% Kai made it to here... (220911 23:40)

\subsubsection{Corrections and systematic errors}
\label{sec:sys}
Since event selection efficiency depends on the PE yield, 
we estimated and applied a correction to compensate for observed PE yield changes particularly during the first one and a half years of data taking. 
%s that are mainly driven by changes in the scintillation light absorption length.
% in the analysis.
This correction and related systematic errors, called relative efficiency and derived from MC simulation, were evaluated and used as described in \cite{XMASS_Modulation, XMASS_Modulation2018}.
%The corresponding signal and background efficiencies %, which depend on the scintillation light absorption length, 
%were evaluated from MC simulation.  
%Simulated background from the PMT window seal's radioactivity, which is the primary background in the low energy region, was used to estimate efficiencies and derive a correction factor for each of the two day periods and its systematic uncertainty.  
To reflect the energy dependence of this relative efficiency for both signal and background events, the energy range from 0.5$-$20 keV$_{\rm ee}$ was divided into four energy bins: 0.5--1~keV$_{\rm ee}$, 1--2~keV$_{\rm ee}$,  2--6~keV$_{\rm ee}$, and 6--20~keV$_{\rm ee}$, and the correction were evaluated separately for each energy bin.
%The mean of 
The resulting mean relative efficiency and its correlated error is shown as a function of time from January 2014 onward in Fig.~\ref{fig:sys}.
%As we
We normalized this relative efficiency and its uncertainty at an absorption length of 8 m for this analysis. %, the relative efficiency, and its correlated error became one and zero at 70 days, respectively.
The mean relative efficiency in the 1--20~keV$_{\rm ee}$ energy range varied from $-$5\% to +10\% for the background events and from about $-$5\% to +4\% for the signal events over the relevant absorption length range. 
Difference between signal and background came from difference of generated position.The majority of background occurred near the detector wall.
The correlated error of this efficiency is the largest  
systematic uncertainty in the analysis.

\begin{figure}[htbp]
\centering
\includegraphics[width=0.52\textwidth]{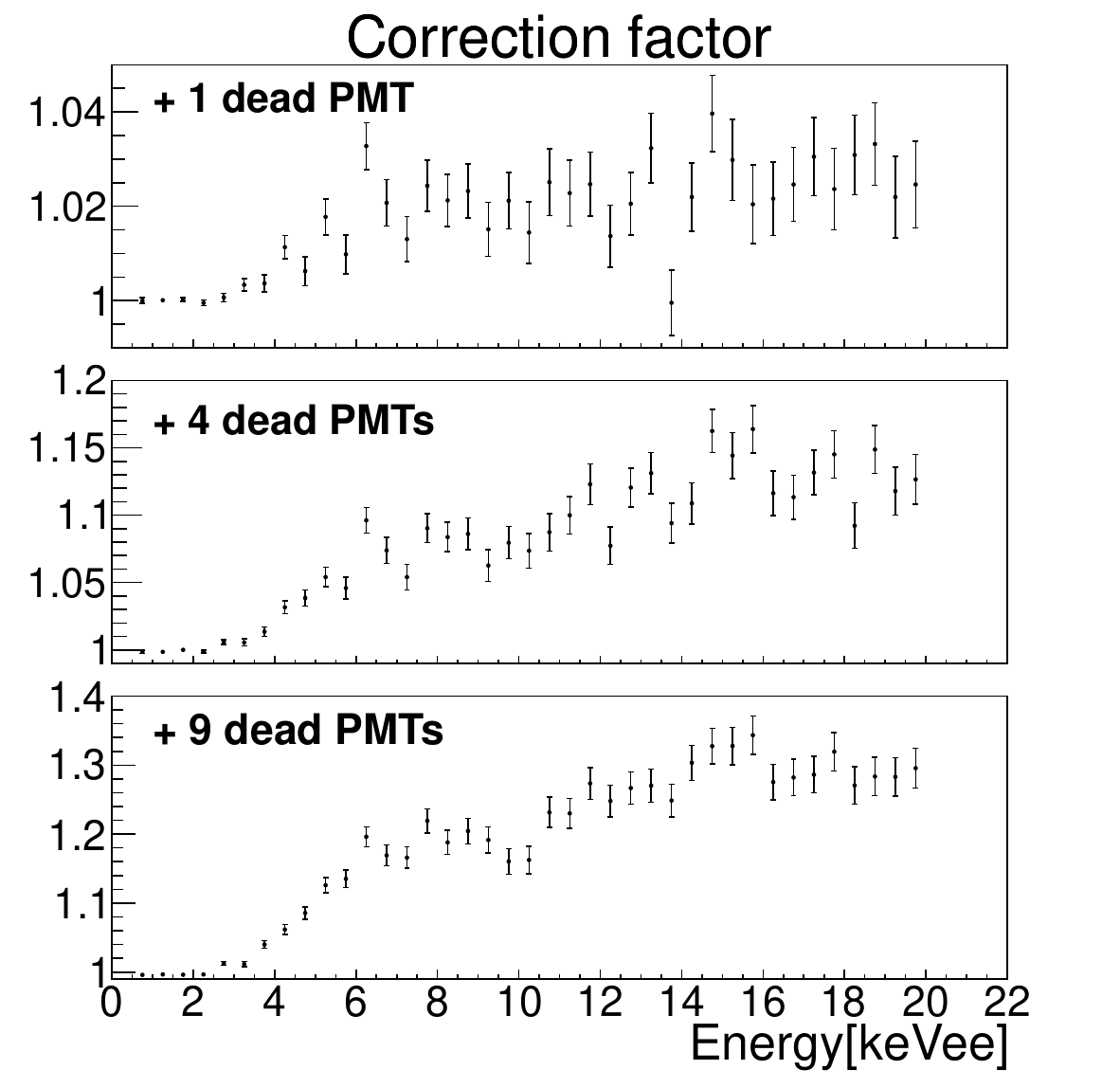}
\caption{Dead PMT correction factors for selection efficiency at + 1, + 4, and + 9 dead PMTs periods. These corrections are applied to correct the selection efficiency from that derived for the 9 dead PMTs period. After that period, number of dead PMT started to increase significantly as shown in Fig.~\ref{fig:dead-pmt}.
Energy dependence of the correction factor comes from differences in the position distribution of the BG events for each energy region.
}
\label{fig:deadeff}
\end{figure}

Subsequently, another correction and its related uncertainty we evaluated to account for change in the number of dead PMTs over time.
The majority of background events occurred in front of the PMT window or near the detector wall.
Since, for such events, a large portion of the emitted scintillation light photons were not registered if the respective PMT was dead, 
event evaluation by the likelihood function were severely affected.
The likelihood function is based on sphericity, aplanarity and the fraction of largest PE. 
As shown in Fig.~\ref{fig:dead-pmt}, the change of the number of dead PMTs was small during the first half of data taking,
but increased from 9 PMTs to 18 in the latter half, making this a significant correction.
The change of selection efficiency with the increase of dead PMTs was estimated using the data taken in the 500 days from May 2015 to September 2016, a time during which the
light yield was stable and the number of dead PMTs did not change, as shown in Fig.~\ref{fig:stability} and \ref{fig:dead-pmt}. 
Deliberately ignoring the signals on selected good PMTs in the analysis, we simulated the dead PMT effect using events from this period and thus estimated
the correction factor to the selection efficiency for each of the subsequently realized dead PMT situations.
The resulting energy dependent correction factors for three subsequently increased numbers of dead PMTs are shown in Fig.~\ref{fig:deadeff}. The errors in the figure are the statistical errors from the data used in the estimation.
Though the method is same as for the FV analysis in Sec.~\ref{sec:deadPMT} to use masking for the effect estimation, the correction in the FV analysis corrects for the difference of the dead tube effect between the observed data and the MC data, 
while the correction here is correcting for the dead tube effect in the observed data. 

During the low-threshold data analysis %, another large 
the uncertainty from the flasher cut against weak light emission,  
explained in Sec.~\ref{Sec:additional}, 
affected the three-hit event selection by at most 0.4\%. 

In addition, between April 2014 and September 2014, 
a gain instability in the waveform digitizers contributed an extra uncertainty of 0.3\% to 
the energy scale.
During that period, a different calibration method was used for the digitizers, causing this instability.
Other uncertainties stemming from the LED gain calibration, trigger-threshold stability, timing calibration, and energy resolution 
were negligible.
  
\subsubsection{Results and Discussion}
\label{sec:NR:results}

For the spin-independent WIMP analysis,  $\chi^{2}$ is defined as:
\begin{equation}
\chi^2 = \sum\limits_{i}\limits^{E_{bins}} \sum\limits_{j}\limits^{t_{bins}} 
\left(
\frac{(R^{{\rm data}}_{i,j}-R^{\rm ex}_{i,j}(\alpha,\beta))^2}
{
\sigma_{{\rm stat}_{i,j}}^2
+\sigma_{{\rm sys}_{i,j}}^2
}
\right)
+\alpha^{2}
%+\sum^{N_{sys}}\beta_i^{2},
+\sum\limits^{N_{sys}}\limits_{i}\beta_i^{2},
\end{equation}
where $R_{i,j}^{\rm data}$, $R_{i,j}^{\rm ex}$,  $\sigma_{{\rm stat}_{i,j}}$, and $\sigma_{{\rm sys}_{i,j}}$ are the data rates and expected MC event rates, and the statistical and systematic errors of the expected event rates for the $i$th energy and $j$th time bin, respectively. The penalty term $\alpha$ relates to the overall size of the relative efficiency error, and it is common for all energy bins; therefore the size of their error simultaneously scales  with $\alpha$ in the fit procedure. $\alpha$=1($-$1) corresponds to the $1\sigma(-1\sigma)$ correlated systematic error as shown in Fig.~\ref{fig:sys} (right) for the expected event rate $R^{\rm ex}_{i,j}(\alpha,\beta)$ in that energy bin. $\alpha$ is determined during $\chi^2$ minimization and increases $\chi^2$ by $\alpha^2$.
The other penalty term, $\beta_i$, relates to the systematic uncertainty of the expected WIMP signal simulation. As explained in~\cite{XMASS_Modulation2019}, this uncertainty has two main components: the scintillation efficiency and the time constant of NR signals. A time constant of 26.9$^{+0.8}_{-1.2}$ ns was used based on our XMASS-I neutron calibration \cite{XMASS_neutron}.
The expected signals are simulated with parameters at the limits of the 1$\sigma$ error range to estimate impacts on the amplitude $A^{s}_{i}(\beta)$ and un-modulated component $C^{s}_{i}(\beta)$ of the respective signal.

The expected modulation amplitudes become a function of the WIMP mass $A_i(m_\chi)$ since the WIMP mass $m_\chi$ determines the recoil energy spectrum.
The expected rate in bin $i,j$ is then proposed, as shown below:
%\begin{equation}
\begin{multline}
R_{i,j}^{\rm ex}(\alpha,\beta) = \int_{t_{j}-\frac{1}{2}\Delta t_{j}}^{t_{j}+\frac{1}{2} \Delta t_{j}} \biggr(\epsilon^b_{i,j}(\alpha)\cdot (B^b_it+C^{b}_{i})\\
 + \sigma_{\chi n} \cdot \epsilon^s_{i,j}\cdot \big( C^{s}_{i}(\beta)+ A^{s}_{i}(\beta) \cos 2\pi \small{\frac{(t-\phi)}{T}} \big) \biggr) dt,
\label{eq:MD}
\end{multline}
where $\phi$ and $T$ are the modulation phase and period, then $t_{j}$ and $\Delta t_{j}$ are the respective time-bin's center and width, $\sigma_{\chi n}$ is the WIMP-nucleon cross section, and $\epsilon^b_{i,j}(\alpha)$ and $\epsilon^s_{i,j}$ are the relative efficiencies for the background and signal, respectively, which are shown in Fig.~\ref{fig:sys} (left).
To account for changing background rates from long-lived isotopes such as $^{60}$Co (t$_{1/2}=$5.27~yr) and $^{210}$Pb (t$_{1/2}=$22.3~yr), we added a linear function with  $B^b_i$ for its slope and $C_i^b$ for its constant term in the $i$th energy bin. $A^s_i(\beta)$ represents an amplitude and $C^{s}_i(\beta)$ a constant for the un-modulated component of the signal in the $i$th energy bin after all cuts.
%at the chosen normalization point on day 70.
The free parameters to be fitted are the cross section $\sigma_{\chi n}$ and  
$B^b_i$ and $C^b_i$ for BG, 
$\alpha$ and $\beta$ are constrained floating parameters.
The observed $R_{i,j}^{\rm data}$ and $\sigma_{{\rm stat}_{i,j}}$ are the input parameters.
To obtain the WIMP-nucleon cross section the data were fitted in the energy range from 0.5 to 20~keV$_{\rm{ee}}$, assuming the same standard halo model as in Sec.~\ref{sec:FV:results}, with the Earth's velocity relative to the DM halo $v_{E} = 232 + 15~{\rm cos} 2\pi(t -\phi)/T$~km/s.
$T$ and $\phi$ were fixed to 365.24 and 152.5 days, respectively.
In this analysis, the signal efficiencies for different WIMP masses were estimated from MC simulations of signal events uniformly distributed in the LXe volume.

\begin{figure}[!t]
  \centering
  \includegraphics[width=9.5cm]{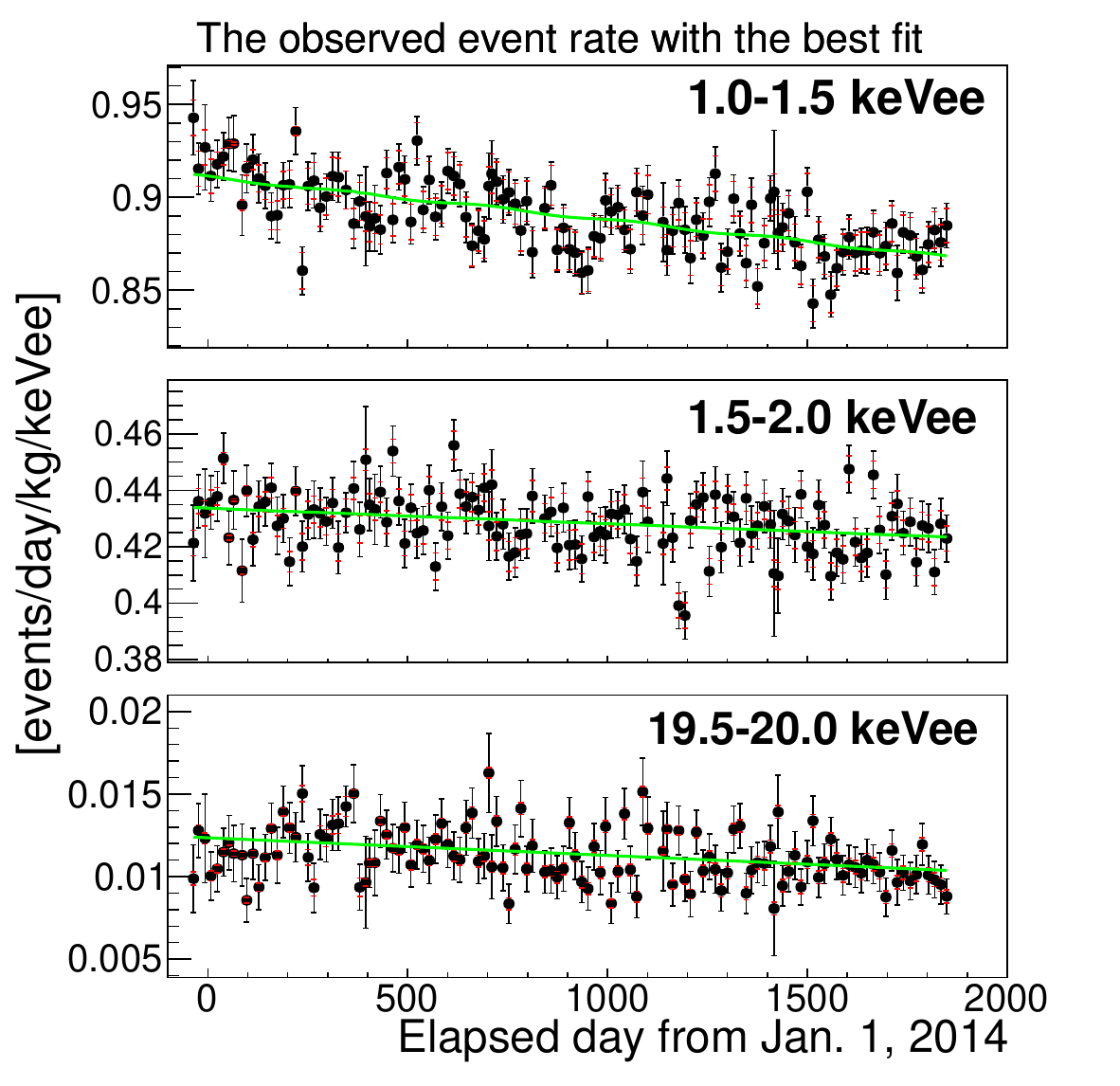}
  \caption{The observed event rate with its best fit and expected time variation for an ${\rm 8\,GeV/c^2}$ WIMP signal within the energy ranges of 1.0--1.5, 1.5--2.0 and 19.5--20.0 keV$_{\rm{ee}}$ as examples of the lowest energy bins and the highest energy bin, respectively.
  The black points indicate data with vertical error bars reflecting the statistical uncertainty of the count rate. 
  The red brackets indicate the 1 $\sigma$ systematic error for each time bin.
  The green line indicates the best-fit result for the 8~${\rm GeV/c^{2}}$ WIMP spectrum 
  with the decaying BG. 
  All data points and lines are corrected for efficiency scaled with the best-fit's $\alpha$. }
  \label{fitting}
\end{figure}

\begin{figure}[b]
\centering
\includegraphics[width=0.5\textwidth]{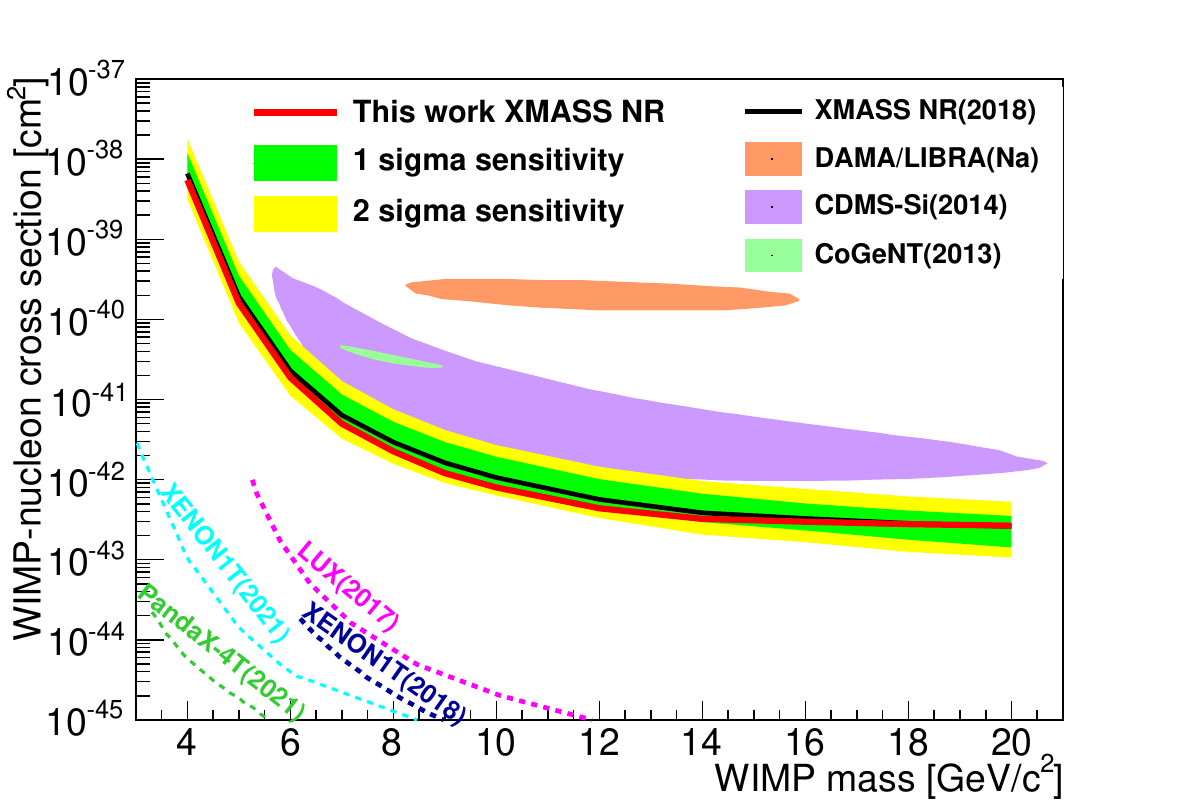}
\caption{Limits on the spin-independent elastic WIMP-nucleon cross section as a function of WIMP mass. The solid red line shows the XMASS 90\% CL exclusion from the annual modulation analysis. The solid black line shows the 2018 XMASS result~\cite{XMASS_Modulation2018}.
The $\pm1\rm{\sigma}$ and $\pm2\rm{\sigma}$ bands represent the expected  90\% exclusion regions. Limits, as well as allowed regions from other searches based on event counting, are also shown~\cite{LUX,XENON1T_1ty,DAMA_LIBRA,CoGeNT_2013,CDMS_Si,XENON1T2021}. }
\label{fig:MD}
\end{figure}
Figure~\ref{fitting} shows the observed event rate with best fit and the expected time variation for ${\rm 8\,GeV/c^2}$ within the energy ranges of 1.0--1.5, 1.5--2.0 and 19.5--20.0 keV$_{\rm{ee}}$.
The best fit for the ${\rm 8\,GeV/c^2}$ WIMP mass had $\chi^2$/ndf =  4806/4734 and $\sigma_{{\chi}n} = (-1.9^{+ 1.6}_{- 2.8}) \times 10^{-42}$ cm$^2$ with the penalty term $\alpha=0.69$.
Since no significant signal was observed, a 90\% CL upper limit on the WIMP-nucleon cross section was set for each WIMP mass, shown in Fig.~\ref{fig:MD} by the red line.
For ${\rm 8\,GeV/c^2}$ WIMP mass, the limit was $2.3 \times 10^{-42}$ cm$^2$.
Here we used the probability function $P$ defined as: 
\begin{equation}
P={\rm exp}\left(-\frac{\chi^2(\sigma_{{\chi}n})-\chi^2_{\rm min} }{2}\right),
\label{eq:prob}
\end{equation}
where $\chi^2(\sigma_{{\chi}n})$ is evaluated as a function of the WIMP-nucleon cross section $\sigma_{{\chi}n}$, and $\chi^2_{\rm min}$ is the minimum $\chi^2$ of the fit.
To obtain our 90\% C.L. exclusion upper limit $\sigma_{up}$, we used the Bayesian approach:

\begin{equation}
 \frac{\int_{0}^{\sigma_{up}} Pd\sigma_{{\chi}n}}{\int^{\infty}_0 Pd\sigma_{{\chi}n}} = 0.9,
\label{eq:integ}
\end{equation}
To evaluate our sensitivity for the WIMP-nucleon cross section, we % carried out a statistical test by 
applied our analysis to 1,000 dummy samples drawn with the same statistical and systematic errors as data, but without any modulation following the procedure described in \cite{XMASS_Modulation}.
The procedure starts by extracting an energy spectrum from the observed data. Then, a toy MC simulation was performed to %produce time variations of background event rates for  each energy bin assuming the same live time as data and including systematic uncertainties.
vary the background event rates in each energy bin incorporating our systematic uncertainty estimates.
The $\pm1 \sigma$ and $\pm2\sigma$ bands in Fig.~\ref{fig:MD} outline the expected 90\% CL upper limit band for the no-modulation hypothesis derived from these dummy samples.

\subsection{%Analysis of
Search for sub-GeV DM}

Conventional xenon detectors are sensitive to DM with sub-GeV masses %\cite{subGeV, subGeV2} 
based on inelastic energy transfer mechanisms involving the Migdal effect or bremsstrahlung photons occurring in NR from collisions involving even very light DM particles~\cite{subGeV, subGeV2}.
%due to the irreducible contribution of the bremsstrahlung effect and Migdal effect accompanying nuclear recoils \cite{subGeV}. 

The bremsstrahlung effect can occur when a DM collision causes a Xe nucleus to accelerate and a bremsstrahlung photon is emitted in the process.
While the energy of such a photon from a DM particle of mass ${\rm 1\,GeV/c^2}$ is limited to 3 keV, its conversion deposits considerably more energy than is transferred in the elastic NR of such light DM particles ($\sim$0.1 keV).
% In the case that a mass of DM particle is ${\rm 1\,GeV/c^2}$, the energy deposited by the bremsstrahlung photon is at most 3 keV. This energy is considerably more than that deposited by elastic nuclear recoil ($\sim$0.1 keV).
The Migdal effect~\cite{Migdal} on the other hand would lead to the emission of an electron from the Xe's atomic shell as the recoiling nucleus accelerates.
Although cross sections for the bremsstrahlung and Migdal effect are smaller than that of elastic NR ($\sim$$10^{-6}$ for Migdal, $\sim$$10^{-8}$ for bremsstrahlung at ${\rm 1\,GeV/c^2}$), the resulting energy deposit becomes much larger due to the inelastic nature of these electromagnetic processes, making it possible to detect recoil even from sub-GeV mass DM particles.
% because these inelastic effects lead to larger energy deposition than elastic nuclear recoil, it should be possible to detect sub-GeV DM through these effects.

\subsubsection{Expected signal}
The expected signal from the Migdal effect was estimated following the %discussion
prescriptions in \cite{Migdal}.
The differential cross section for this process %with respect to
as a function of the %atomic
NR energy is:

\begin{equation}
\begin{split}
\label{eq:dsigder}
 \frac{d\sigma}{dE_R} \simeq
 \sum_{E_{ec}^{F}}
 \frac{1}{32\pi}\frac{m_A}{\mu^2_{N}v^2_{DM}}
  \frac{|F_A(q^2_A)|^2|
  \mathcal{M}(q_A)|^2}{(m_A + m_{DM})^2}|Z_{FI}(q_e)|^2,
\end{split}
\end{equation}
where $m_A$ is the physical mass of the atomic system including the electron cloud energy
%defined by 
${m_A \simeq m_N + N_em_e + E_{ec}}$, 
the DM particle mass $m_{DM}$, % is the dark matter mass,
the DM-nucleon reduced mass $\mu_N$, % is the DM-nucleon reduced mass, 
the DM particle velocity in the laboratory frame $v_{DM}$, % is the dark matter velocity in the laboratory frame,
the nuclear form factor $F_A$ for momentum transfer $q_A$, %is the nulclear from factor which is relevant for a momentum transfer $q_A$,
the invariant amplitude $\mathcal{M}$, % is the invariant amplitude,
and a factor $Z_{FI}(q_e)$ capturing the transition probability in the electron cloud with  %denoting the transition of the electron cloud, 
$E_R \simeq \frac{q^2_A}{2m_A}$ and
$q_e \simeq \frac{m_e}{m_A}q_A$.

The %dark matter
Migdal event rate per unit detector mass and time is given by:

\begin{multline}
\label{eq:eventrateER}
\frac{dR}{dE_Rdv_{DM}} \simeq \sum_{E_{ec}^{F}}\frac{1}{2}\frac{\rho_{DM}}{m_{DM}}\frac{1}{{\mu}^2_N}
\tilde{\sigma}_N(q_A)\\
\times|Z_{FI}(q_e)|^2
\times\frac{\tilde{f}(v_{DM})}{v_{DM}},
\end{multline}
where
\begin{equation}
\begin{split}
\label{eq:tildesigma}
\tilde{\sigma}_N(q_A) = \frac{1}{16\pi}\frac{|F_A(q^2_A)|^2|\mathcal{M}(q^2_A)|^2}{(m_A+m_{DM})^2},
\end{split}
\end{equation}
with $\rho_{DM}$ denoting the local DM density and $\tilde{f}(v_{DM})$ the DM particle velocity 
distribution integrated over all directions. %the directional component.

% Then the energy spectrum of the electron by the Migdal effect from the initial orbit (n,l) is 
The energy spectrum for Migdal emission from an initial orbit (n,l) then becomes
\begin{equation}
    \begin{split}
\label{eq:eventrateEe}
\frac{dR}{dE_RdE_edv_{DM}} \simeq \frac{dR_0}{dE_Rdv_{DM}}\times\frac{1}{2\pi}\sum_{n,l}
\frac{d}{dE_e}p^c_{q_e}(nl\rightarrow E_e),
    \end{split}
\end{equation}
with
\begin{eqnarray}
\label{eq:dR0}
\frac{dR_0}{dE_Rdv_{DM}} \simeq \frac{1}{2}\frac{\rho_{DM}}{m_{DM}}\frac{1}{{\mu}^2_N}
\tilde{\sigma}_N(q_A)\times\frac{\tilde{f}(v_{DM})}{v_{DM}},
\end{eqnarray}
if $p^c_{q_{e}}$ is the ionization probability.
When calculating the expected signal in the XMASS detector 
the energy dependent scintillation light yield is calculated separately for electron emission from the inner shell and the subsequent de-excitation emission. 
% The respective numbers of scintillation photons were calculated separately for ionized electron and de-excitation not for total energy loss by both. 

The annual modulation of the bremsstrahlung signal is evaluated in the same way as in \cite{XMASS_Modulation2019}.
The corresponding differential event rate is:
\begin{equation}
\label{eq:velavg}
 \frac{dR}{d\omega} = N_T \frac{\rho_{DM}}{m_{DM}} \int_{ v\geq v_{\rm min}} 
 d^3 v v f_v({\bf v} + {\bf v_E}) \frac{d\sigma}{d\omega} ,
\end{equation}
where $N_T$ is the number of target nuclei per unit mass in the detector,  
${\bf v_{E}}$ is the velocity of the Earth relative to the galactic rest frame, 
and $f_v({\bf v})$ is the DM velocity distribution in the galactic frame. 
The minimum velocity was $v_{\rm min}=\sqrt{2\omega/\mu_N}$ \cite{LewinSmith}.
We used the same parameters as in our prior multi-GeV analysis in Sec.~\ref{sec:NR:results}.

\begin{figure}[!t]
  \centering
  \includegraphics[width=9.4cm]{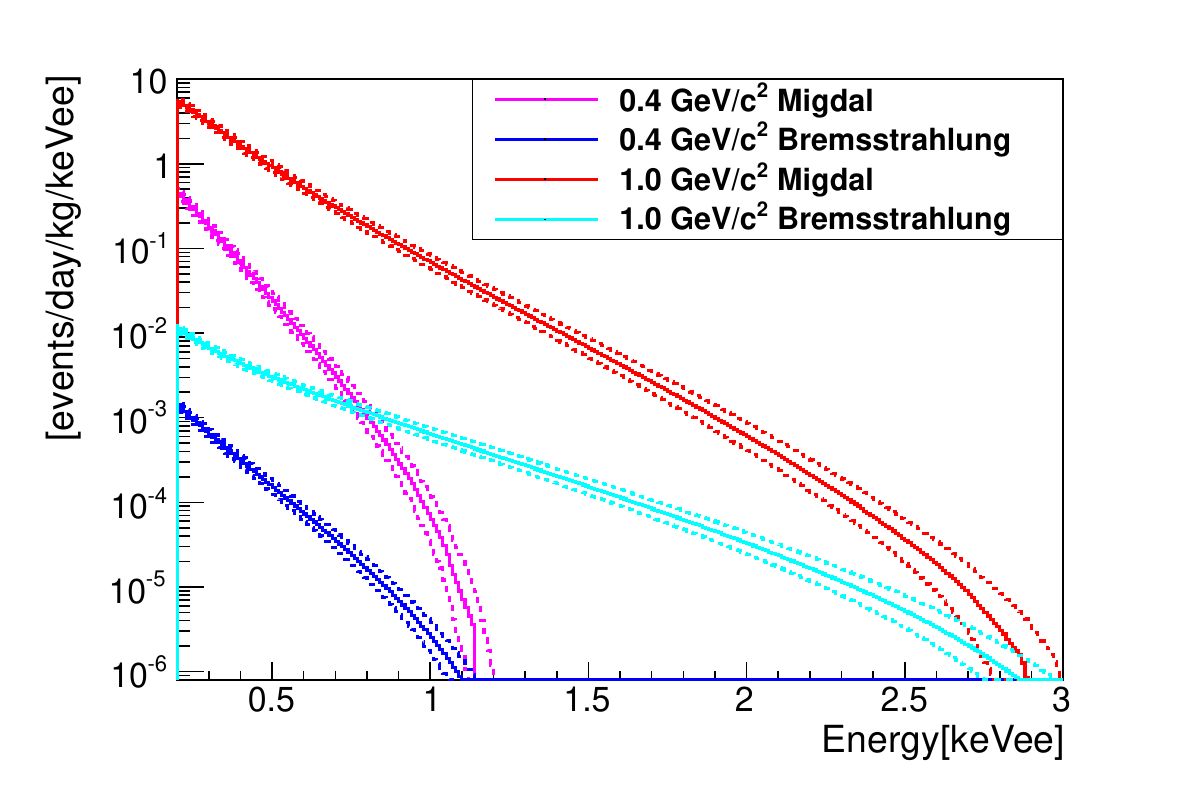}
  \caption{Expected energy spectra from Migdal effect and bremsstrahlung caused by ${\rm 0.4\,GeV/c^2}$ WIMPs (magenta for Migdal and cyan for bremsstrahlung) and ${\rm 1.0\,GeV/c^2}$ WIMPs (red for Migdal, blue for Bremasstrahlung) with a cross section of $10^{-35}\,{\rm cm}^{2}$. 
The dotted line shows the respective spectra in June and December, while the solid one represents the annual average spectrum. 
No detector effects are considered yet in these spectra.}
  \label{signal}
\end{figure}

%We also calculated the event rate as a function of bremsstrahlung energy and time.
Figure \ref{signal} shows the expected Migdal and bremsstrahlung spectra for ${\rm 0.4\,GeV/c^2}$ DM interactions in June and December corresponding to the maximum and minimum relative velocity $v_{E}$, respectively, as well as the average spectrum.
The resulting expected annual modulation amplitude was about 30\% of the average event rate at 1 keV before considering detector effects such as energy non-linearity and resolution.

Since in this energy region the signal from NR alone is negligibly small compared to that from the Migdal effect and bremsstrahlung, 
a NR contribution was not considered in these analyses.

\subsubsection{Results and discussion}

For the sub-GeV DM analysis almost the same fitting procedure as discussed in Sec.~\ref{sec:NR:results} was applied. 
Differences stem from the fundamental ER nature of these signals. %differing mechanisms %related
% producing these signals.
The expected signal rates were estimated for both Migdal effect and bremsstrahlung, with the 
uncertainties in the relevant scintillation decay constants and scintillation efficiency for ER signals properly considered.
These uncertainties introduce correlations between energy bins in the signal spectrum.
For the scintillation decay time constants, two components, referred to as the fast and the slow component, were used, based on our XMASS-I $\gamma$-ray calibrations~\cite{XMASS_decayt}.
These were 2.2 ns and 27.8$^{+1.5}_{-1.0}$ ns, respectively, with the fast component's fractional contribution at 0.145$^{+0.022}_{-0.020}$.

%The analysis was conducted
Signal spectra were calculated for DM masses from ${\rm 0.32}$ to ${\rm 1\,GeV/c^2}$ for bremsstrahlung %effect
and from ${\rm 0.35}$ to ${\rm 4\,GeV/c^2}$ for Migdal mediated signals.
The lower limits of these mass ranges were determined by the requirement to deposit more than 1 keV$_{\rm{ee}}$ %signal
in the detector. Below that DM mass, the expected number of events which deposit more than 1 keV$_{\rm{ee}}$ decrease sharply.
The higher limit of ${\rm 1\,GeV/c^2}$ for bremsstrahlung is same as used in \cite{XMASS_Modulation2019}, and is based on the same assumptions as made for the signal calculation in
\cite{subGeV}. %, such as that for form factor were not proper. (Kai: ??? assumptions for the form factor were not proper ??? did not understand, but also think the reference to the old analysis frees us from repeating the argument...
The upper limit of ${\rm 4\,GeV/c^2}$ seems reasonable as beyond this energy the sensitivity of the conventional NR analysis becomes much higher than that of the bremsstrahlung and Migdal analyses.
 
\begin{figure}[h!]
  \centering
  \includegraphics[width=9cm]{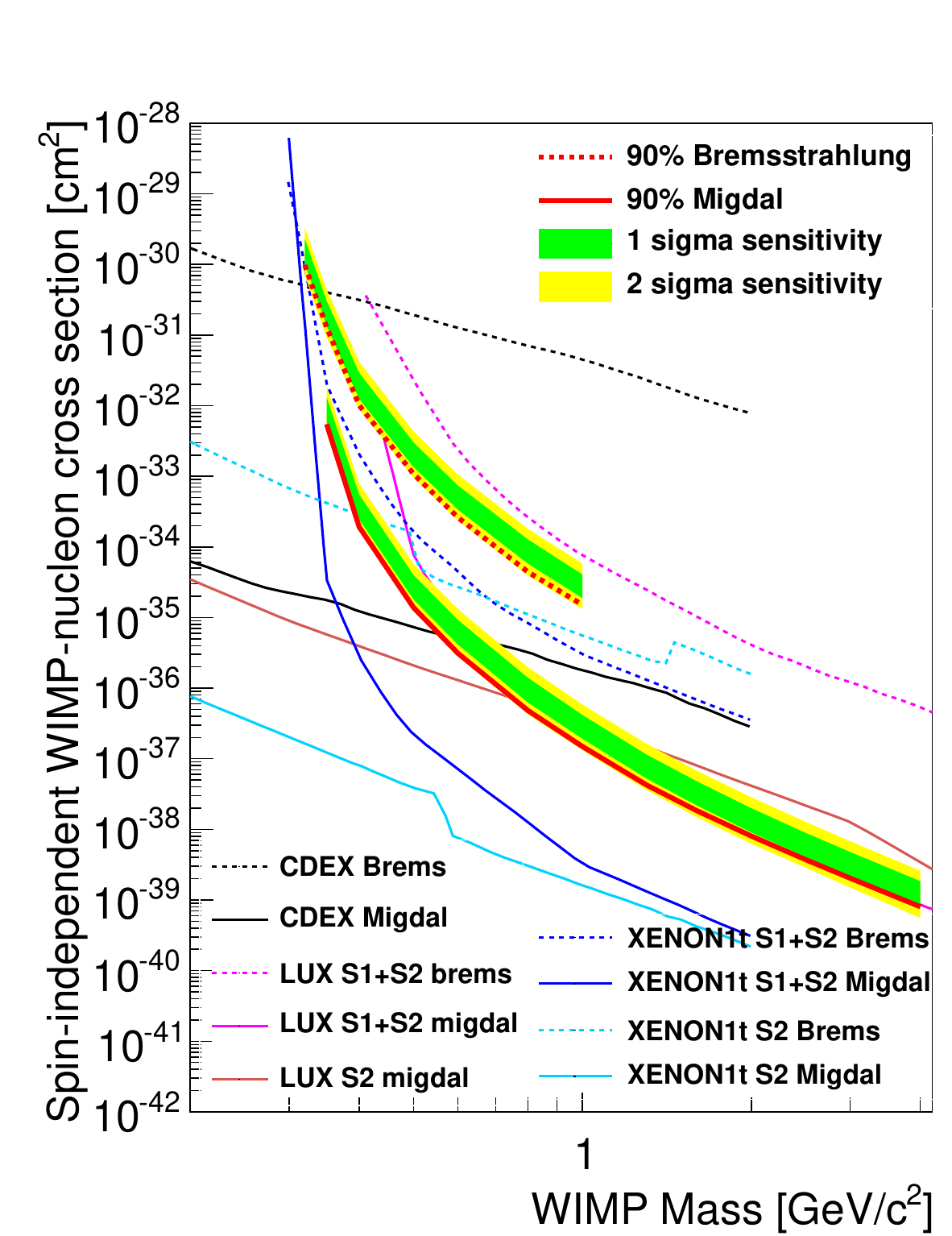}
  \caption{
  Summary of the sub-GeV DM analysis results considering annual modulation with the Migdal effenct and the bremsstrahlung. The red solid line is the result of the Migdal analysis for 0.35-4$\,{\rm GeV/c^2}$ WIMP mass, the red dotted line is the result of the bremsstrahlung analysis for 0.32-1$\,{\rm GeV/c^2}$. Limits from other experiments obtained via Migdal and Bremsstrahlung effect are also shown for comparison~\cite{XENON1T_Migdal,LUX_S1S2Migdal,LUX_S2Migdal,CDEX}.
}
  %The red line is the result of the bremsstrahlung analysis for 0.32--1 GeV DM.
  %%This is the first result using modulation and bremsstrahlung effects for sub-GeV DM.
  %For comparison, data from the CRESST sapphire surface detector \cite{CRESSTSurPaper} and %CRESST-II \cite{CRESSTIIPaper}, which are searching for the elastic nuclear recoil signals, are %shown in each colour.
  %The black line shows the result of the nuclear recoil search at 4--20 GeV.
  %For comparison, results from CDMS-Si \cite{CDMSSiPaper}, CDMSLite \cite{CDMSLitePaper}, SuperCDMS \cite{SuperCDMSPaper}, \textcolor{red}{LUX (NR \cite{LUX}, Bremsstrahlung and Migdal \cite{LUX_Brems_Migdal})} , XENON1T \cite{XENON1T}, PandaX-II \cite{PANDA}, DAMA/LIBRA \cite{DAMAPaper, DAMA_WIMP}, and XMASS-I \cite{XMASS_MOD2017}, DarkSide-50 \cite{DS}, and the liquid scintillator experiment by Collar \cite{JIC} are shown for each colour.
  %The green and yellow bands for each result show the $\pm$ 1 $\sigma$ and $\pm$ 2 $\sigma$ %expected sensitivity of 90\% CL upper limits for the null-amplitude case, respectively.
  
  \label{Summary}
\end{figure}

The best fit cross section from our data was $(-2.8^{+1.5}_{-2.0}) \times 10^{-35}$ cm$^2$ at ${\rm 0.5\,GeV/c^2}$ for the Migdal analysis %effect. 
with a $\chi^2$/ndf of 4739/4670, and the penalty term $\alpha$ becoming ${\rm 0.67}$.
%The calculation of sensitivities for the null-amplitude case will be discussed below.
The result of the DM searches via Migdal and bremsstrahlung effects in the sub-GeV WIMP mass region is shown in Fig.~\ref{Summary}.
The expected sensitivity for the null-amplitude case was again calculated using toy MC samples. %the statistical samples.
The 90\% CL sensitivity for DM at ${\rm 0.5\,GeV/c^2}$ was $(2.7^{+1.3}_{-0.9}) \times 10^{-35}$ cm$^2$ (the range containing 68\% of the toy MC samples) for the Migdal analysis and $(2.0^{+1.0}_{-0.6}) \times 10^{-33}$ cm$^2$ for the bremsstrahlung analysis.
Our upper limits with a p-value of 0.09 for ${\rm 0.5\,GeV/c^2}$ were 1.38 $\times 10^{-35}$ cm$^2$ for Migdal and 1.1 $\times 10^{-33}$ for bremsstrahlung.

%{\bf [very hard to see the legend in Fig.20 ]}
%{\bf [explain why we analyzed up to 1 GeV and not more. ]}
%{\bf [explain why we dont' have the curve btw 1 and 4 GeV in Fig22. ]}

\subsection{Model-independent analysis}

 For the model-independent analysis, our $\chi^2$ %is defined as:
 was defined, as shown below.
 
\begin{equation}
\chi^2 = \sum\limits_{i}\limits^{E_{bins}} \sum\limits_{j}\limits^{t_{bins}} 
\left(\frac{(R^{{\rm data}}_{i,j}-R^{\rm ex}_{i,j})^2}{\sigma({\rm stat})^2_{i,j}+\sigma({\rm sys})^2_{i,j}}\right)+\alpha^{2}, 
\end{equation}
with the expected event rate being
%\begin{equation}
\begin{multline}
%\begin{split}
%R_{i,j}^{\rm ex} = \int_{t_{j}-\frac{1}{2}\Delta t_{j}}^{t_{j}+\frac{1}{2} \Delta t_{j}} \left( C_{i} +A_{i} \cos 2\pi \small{\frac{(t-t_{0})}{T}} \right) dt, \\
R_{i,j}^{\rm ex} = \int_{t_{j}-\frac{1}{2}\Delta t_{j}}^{t_{j}+\frac{1}{2} \Delta t_{j}} 
\left( \epsilon^s_{i,j} A^s_{i} \cos 2\pi \small{\frac{(t-\phi)}{T}} \right.\\
\left. + \epsilon^b_{i,j}(\alpha) (B^{b}_it + C^{b}_{i})\right) dt,
\label{eq:MI}
%\end{split}
\end{multline}
%\end{equation}
where $C_{i}^b$ and $A_{i}^s$ are free parameters for the unmodulated event rate and the modulation amplitude without absolute efficiency correction, respectively. In the fitting procedure, the energy range 1--20~keV$_{\rm ee}$ was used, the modulation period $T$ was fixed to one year (= 365.24~days), and the phase %pinned to
$\phi$ was fixed to 152.5 days ($\sim$2nd of June), the time when the Earth's velocity relative to the DM distribution is expected to be maximal. 
%The observed count rate after cuts as a function of time in the energy %region between 1.0 and 3.0~keV$_{\rm{ee}}$ is shown in %Fig.~\ref{fig:rate}. For an easy visualization, the data points were %corrected for relative efficiency based on the best-fit parameters %instead of  the fitting function, therefore, the fitted line in %Fig.~\ref{fig:rate} is simply a cosine plus a one-dimensional %polynomial function. 

%\begin{figure}[tp]
%\centering
%\includegraphics[width=0.5\textwidth]{fig7.pdf}
%\caption{Observed count rate as a function of time in the 1.0 -- 3.0 %keV$_{\rm ee}$  energy range after correcting relative efficiency  (see %text).  The black error bars show the statistical uncertainty of the %count rate. The solid curves represent  the best fit result for a %model-independent analysis before correcting for total efficiency.}
%\label{fig:rate}
%\end{figure}

\begin{figure}[tb]
\centering
\includegraphics[width=0.52\textwidth]{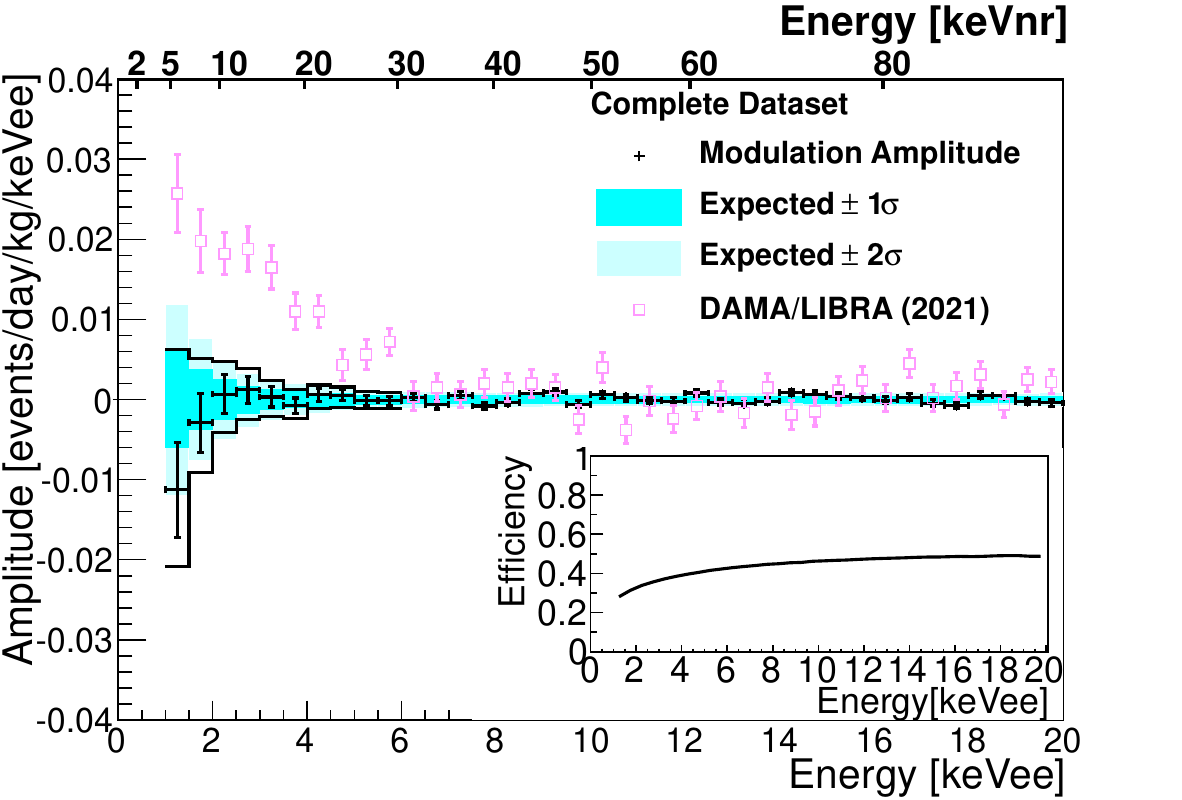}
\caption{Modulation amplitude as a function of energy for the model-independent analysis using the pull method (black cross). Solid lines represent the 90\% positive (negative) upper limits on the amplitude. The $\pm1\rm{\sigma}$ and $\pm2\rm{\sigma}$ bands represent the expected amplitude % region
fluctuation ranges for a null result (for details, see text). 
The small figure at bottom right shows
the signal efficiency used in the model-independent analysis, 
the overall efficiency for a uniformly distributed electron signal after applying all selection criteria.
The latest DAMA/LIBRA result (square) is shown for reference~\cite{dama2022}. }
\label{fig:MI}
\end{figure}

 The best-fit in the energy region between 1 and 20~keV$_{\rm{ee}}$ for our modulation hypothesis, with the fixed phase and period as detailed above, yielded $\chi^2_1$/ndf = 4693/4635 with $\alpha=0.74\pm0.04$. The result for the null hypothesis (fixing  $A^s_i = 0 $) was $\chi^2_0$/ndf  = 4741/4673 with $\alpha=0.74\pm0.08$.
Figure~\ref{fig:MI} shows the best-fit amplitudes as a function of energy after correcting for efficiency using the curve showed at bottom right in the same figure.  The $\pm1\rm{\sigma}$ and $\pm2\rm{\sigma}$ bands in Fig.~\ref{fig:MI} represent our expected amplitude coverage derived again from the same dummy sample procedure as in the analyses above.
A hypothesis test was also done with these dummy samples, using their $\chi^2$ difference $\chi^2_0-\chi^2_1$ to obtain a $p$-value
%This test gave the $p$-value
of 0.14 (1.5$\sigma$) for this best-fit result.
  
% To be able to test any model of DM, 
Not to limit the models that can be checked against our data we evaluated the constraints on the positive and negative amplitudes separately in Fig.~\ref{fig:MI}.
The upper limits on the amplitudes in each energy bin were calculated considering separately the regions of positive or negative amplitudes by integrating Gaussian distributions based on the mean and sigma of the data (=$G(a)$) from zero. The positive or negative upper limits were %satisfied 
derived with 0.9 for $\int_0^{a_{up}} G(a)da/\int_0^\infty G(a)da$ or $\int_{a_{up}}^{0} G(a)da/\int_{-\infty}^0 G(a)da$, where $a$ and $a_{up}$ are the amplitude and its 90\% CL upper limit, respectively.
This method obtained a positive (negative) upper limit of 
% $0.96~(-1.5)\times10^{-2}$
$0.62~(-2.1)\times10^{-2}$ events/day/kg/keV$_{\rm{ee}}$ between 1.0 and 1.5 keV$_{\rm{ee}}$ with the limits becoming stricter at higher energy. The energy resolution ($\sigma/$E) at 1.0(5.0) keV$_{\rm{ee}}$ was estimated to be 36\% (19\%), comparing our gamma ray calibration data to our MC simulation.
%As a guideline, we make the direct comparisons with other experiments not by considering a specific dark matter model.
A modulation amplitude of $\sim2 \times10^{-2}$ events/day/kg/keV$_{\rm ee}$ was obtained by DAMA/LIBRA between 1.0 and 3.5 keV$_{\rm ee}$~\cite{dama2022}, while our positive upper limit was  $\sim5 \times10^{-3}$ events/day/kg/keV$_{\rm{ee}}$ in that same energy range.
%, and  XENON100 reported $(1.67\pm0.73)\times10^{-3}$ events/day/kg/keV$_{\rm ee}$ (2.0$-$5.8  keV$_{\rm ee}$) \cite{XENON_MOD}. This result corresponds to a 90\% CL upper limit (one-sided) of $2.9\times10^{-3}$ events/day/kg/keV$_{\rm ee}$.  Our study obtained a 90\%~CL positive upper limits of $(1.3-3.2)\times10^{-3}$ events/day/kg/keV$_{\rm ee}$ in the same energy region  and  gives the more stringent constraint above 3.0 keV$_{\rm ee}$  as shown in Fig.\ref{fig:MI}. This fact is important when we test dark matter model.

\section{Conclusions}
\label{Sec:Conclusions}

\begin{figure}[t]
\centering
\includegraphics[width=0.5\textwidth]{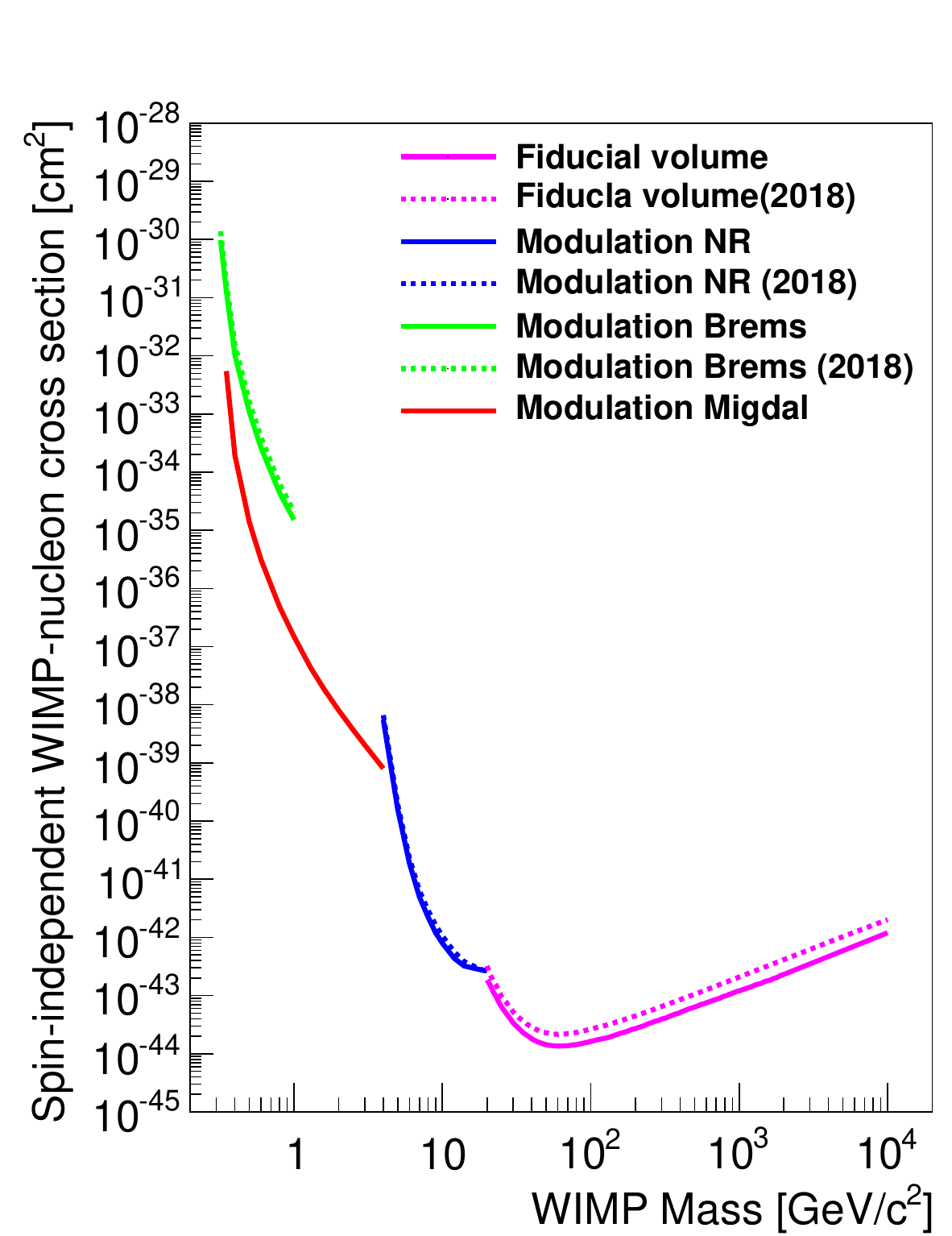}
\caption{Limits obtained from the various analyses presented in this paper. Solid lines are the new results, using the full XMASS-I data sets, presented in this paper, in particular the fiducial volume analysis, the muliti-GeV modulation analysis via NR and the sub-GeV modulation analysis via Migdal effect and bremsstrahlung.  Dotted lines show our earlier results published in \cite{XMASS_FV,XMASS_Modulation2018,XMASS_Modulation2019}}.
\label{fig:compile}
\end{figure}

%As unique single phase liquid Xenon detector, XMASS-I took over 5 years data.
XMASS-I was a unique single-phase LXe detector, which took data almost continuously over 5 full years.
%Quite
Over this long period of stable observation it accumulated 1590.9 live days of data with an analysis threshold of 1~${\rm keV_{ee}}$.
A subset of 768.8 days therein allows for an even lower analysis threshold of 0.5~${\rm keV_{ee}}$.
%for normal threshold trigger data (1${\rm keV_{ee}}$ threshold) and low threshold trigger data (0.5${\rm keV_{ee}}$).

Extending the FV search with a target mass of 97 kg to the full 1590.9 days allowed us to improve our earlier world-best single phase LXe limit on spin-independent high mass WIMP interactions by a factor of 1.6 down to ${\rm 1.4\times 10^{-44}\, cm^{2}}$ for a ${\rm 60\,GeV/c^{2}}$ WIMP at the 90$\%$ CL.
% By using this full set we carried out two WIMP dark matter searches.
% One is the WIMP signal search with fiducial volume selection. 
% By fitting the energy spectrum after all selection with expected signal and BG, obtained 
% upper limit on a spin-independent WIMP-nucleon cross section was ${\rm 1.4\times 10^{-44}\, cm^{2}}$ for a ${\rm 60\,GeV/c^{2}}$ WIMP ${\rm 2.2\,GeV/c^{2}}$  at the 90$\%$ confidence level. This is factor 1.6 better results 
% rom previous XMASS-I results.

Updated searches for an annual modulation signature expected for true galactic DM halo particle interactions in terrestrial detectors, % pushed our unique single-phase LXe limits, 
now also extended to the full XMASS-I data set and using XMASS-I's full active target mass of 832 kg, improved on our own old limits for NR by a factor of 1.3 to reach 
${\rm 2.3\times 10^{-42}\, cm^{2}}$ for a ${\rm 8\,GeV/c^{2}}$ WIMP.
Also updated was our bremsstrahlung result, which 
% Our bremsstahlung based result also updated.
%As a typical number, the result 
for a ${\rm 0.5\,GeV/c^{2}}$ WIMP %got updated 
now reached a cross section limit of ${\rm 1.1\times 10^{-33}\, cm^{2}}$, an improvement of a factor 1.5.
The newly added analysis exploiting the Migdal effect for low mass WIMP searches closed the WIMP mass gap that previously existed in our analyses between the lower WIMP mass end of the NR modulation analysis and the upper WIMP mass end of our bremsstrahlung based modulation analysis, reaching down to ${\rm 1.4\times 10^{-35}\, cm^{2}}$ for ${\rm 0.5\,GeV/c^{2}}$ WIMPs.

Altogether, as summarized in Fig.~\ref{fig:compile}, XMASS-I WIMP searches cover the whole mass range from 0.32 to $10^{4}$~GeV/$c^{2}$ with our cross section limits in a single detector,
and are the world best results from a single phase LXe detector.

\section*{Acknowledgements}
We gratefully acknowledge the cooperation of the Kamioka Mining and
Smelting Company. This work was supported by the Japanese Ministry of
Education, Culture, Sports, Science and Technology, Grant-in-Aid for Scientific Research, JSPS KAKENHI Grant No. 19GS0204, 26104004,
and 19H05805, the joint research program of the Institute for Cosmic Ray
Research (ICRR), the University of Tokyo, and partially by the National Research Foundation of Korea Grant funded by the Korean Government (NRF2011-220-C00006), and the Brain Korea 21 FOUR Project grant funded by
the Korean Ministry of Education.


\section*{References}
\begin{thebibliography}{00}
\bibitem{Faber} S.~M. Faber and J.~S. Gallagher, Annu. Rev. Astron. Astrophys. 17 (1979) 135.
\bibitem{Beringer} C. Patrignani \textit{et al.}, Particle Data Group, Chin. Phys. C 40 (2016) 010001.
%\bibitem{XMASS_det} K. Abe \textit{et al.}, (XMASS Collaboration), Nucl. Instr. and Meth. A716 (2013) 78.
\bibitem{Goodman} M.~W. Goodman, E. Witten, Phys. Rev. D 31 (1985) 3059.
\bibitem{Gondolo} Paolo Gondolo arXiv:hep-ph/9605290.
%\bibitem{PandaX} Xiangyi Cui, et al., PandaX-II Collaboration, Phys. Rev. Lett. 119 (2017) 181392.
%\bibitem{XENON1T} E. Aprile, et al., XENON Collaboration, Phys. Rev. Lett. 119 (2017) 181301.
%\bibitem{LZ2022} J. Aalbers \textit{et al.} (LUX-ZEPLIN (LZ) Collaboration), arXiv:hep-ex/2207.03764v2.
\bibitem{XENON1T_1ty} E. Aprile \textit{et al.} (XENON Collaboration), Phys. Rev. Lett. 121 (2018) 111302.
\bibitem{XENON1T_Migdal} E. Aprile \textit{et al.} (XENON Collaboration), Phys. Rev. Lett. 123 (2019) 241803.
\bibitem{LUX} D.~S. Akerib \textit{et al.} (LUX Collaboration), Phys. Rev. Lett. 118 (2017) 021303.
%\bibitem{PandaX-II_full} Qiuhong Wang et al., (PandaX-II Collaboration) Chinese Phys. C 44 (2020)125001
\bibitem{PandaX-4T} Yue Meng \textit{et al.} (PandaX-4T Collaboration), Phys. Rev. Lett. 127 (2021) 261802.
\bibitem{DARKSide} P. Agnes \textit{et al.} (DarkSide Collaboration), Phys. Rev. D 98, 102006 (2018). %arXiv:1802.07198v1.
\bibitem{DEAP3600} R. Ajaj \textit{et al.} (DEAP Collaboration), Phys. Rev. D 100, 022004 (2019).
\bibitem{XMASS_det} K. Abe \textit{et al.}, (XMASS Collaboration), Nucl. Instr. and Meth. A716 (2013) 78.
%Phys. Rev. Lett. 121, 071801 (2018)
%arXiv:1707.08042v2.
%%\bibitem{CRESST} G. Angloher, et al., CRESST Collaboration, Eur. Phys. J. C (2012) 72:1971
\bibitem{XMASS_LowMassWIMP} K. Abe \textit{et al.} (XMASS Collaboration), Phys. Lett. B 719 (2013) 78.
\bibitem{XMASS_inelastic} U. Uchida \textit{et al.} (XMASS Collaboration), Prog. Theor. Exp. Phys. (2014) 063C01.
\bibitem{XMASS_Modulation} K. Abe \textit{et al.} (XMASS Collaboration), Phys. Lett. B 759 (2016) 272.
\bibitem{XMASS_bosonic} K. Abe \textit{et al.} (XMASS Collaboration), Phys Rev Lett, 113, (2014) 121301.
\bibitem{XMASS_FV} K. Abe \textit{et al.} (XMASS Collaboration), Phys. Lett. B 789 (2019) 45-53.
\bibitem{XMASS_Modulation2018} K. Abe \textit{et al.} (XMASS Collaboration), Phys. Rev. D 97, 102006 (2018).
\bibitem{XMASS_Modulation2019} M. Kobayashi \textit{et al.} (XMASS Collaboration), Phys.Lett. B 795 (2019) 308.
%\bibitem{Migdal} M.~Ibe, J. High Energ. Phys, 2018 194 (2018).
\bibitem{XMASS_inelastic2019} K. Abe \textit{et al.} (XMASS Collaboration),  Astropart. Phys. 110 (2019) 1–7.
\bibitem{XMASS_AXION2013}
K. Abe \textit{et al.} (XMASS Collaboration), Phys.Lett. B724 (2013) 46-50.
\bibitem{XMASS_KK} N. Oka \textit{et al.} (XMASS Collaboration), Prog. Theor. Exp. Phys. 10 (2017), 103C01.
\bibitem{XMASS_hiddenphoton} K. Abe \textit{et al.} (XMASS Collaboration), Phys. Lett. B787 (2018) 153-158.
\bibitem{XMASS_DEC} K. Abe \textit{et al.} (XMASS Collaboration),
 Phys. Lett. B759 (2016) 64-68.
\bibitem{XMASS_DEC2018} K. Abe \textit{et al.} (MASS Collaboration), Prog. Theor. Exp. Phys. 2018 (2018) 053D03.
\bibitem{XMASS_SN}K. Abe \textit{et al.} (XMASS Collaboration), 
Astropart. Phys. 89 (2017) 51-56.
\bibitem{XMASS_exotic} K. Abe \textit{et al.} (XMASS Collaboration), Phys. Lett. B 809 (2020) 135741.
\bibitem{XMASS_GW} K. Abe \textit{et al} (XMASS Collaboration),  Astropart. Phys. 129 (2021) 102568.
\bibitem{XMASS_R10789}
K. Abe \textit{et al.} (XMASS Collaboration), Nucl. Instrum. and Meth. A922 (2019) 171-176. 
\bibitem{SK_ATM} H. Ikeda \textit{et al.}, Nucl. Instrum. and Meth A 320 (1992) 310.
S. Fukuda \textit{et al.} (Super-Kamiokande Collaboration), Nucl. Instrum. and Meth. A 501 (2003) 418.
\bibitem{XMASS_cal} N.Y. Kim, \textit{et al.} (XMASS Collaboration), Nucl. Instr. and Meth. A784 (2015) 499.
\bibitem{dpe1} C.~H. Faham \textit{et al.}, JINST 10 (2015) P09010.
\bibitem{dpe2} B. López Paredes \textit{et al.}, Astropart. Phys. 102 (2018) 56–66.
%\bibitem{XENON2011} E. Aprile \textit{et al.} (The XENON100 Collaboration), Phys. Rev. Lett. 107 (2011) 131302.
\bibitem{XMASS_neutron} K. Abe \textit{et al}. (XMASS Collaboration), JINST 13 (2018) P12032.
%\bibitem{XMASS_cal} N.Y. Kim, \textit{et al.} (XMASS Collaboration), Nucl. Instr. and Meth. A784 (2015) 499.
%\bibitem{XMASS_takeda} A. Takeda for the XMASS Collaboration, Proceedings of 34th International Cosmic Ray Conference, PoS (ICRC2015) 1222. 
\bibitem{SK_PCRflux} G. Guillian \textit{et al.} (Super-Kamiokande Collaboration), Phys. Rev. D 75, 062003 (2007).
\bibitem{XMASS_RT} A. Takeda for the XMASS Collaboration, Proceedings of 34th International Cosmic Ray Conference, PoS (ICRC2015) 1222.
%\bibitem{XMASS_neutron} K. Abe \textit{et al}. (XMASS Collaboration), JINST 13 (2018) P12032.
\bibitem{XMASS_HPGe} K. Abe \textit{et al}. (XMASS Collaboration), Nucl. Instrum. Meth. A922 (2019) 171-176.
\bibitem{XMASS_alpha} K. Abe \textit{et al}. (XMASS Collaboration), Nucl. Instrum. Meth. A884 (2018) 157-161.
\bibitem{LewinSmith2} M.~C. Smith \textit{et al.}, Mon. Not. R. Astron. Soc. 379 (2007) 755.
\bibitem{LewinSmith} J.~D. Lewin, P.~F. Smith, Astropart. Phys. 6 (1996) 87.
\bibitem{SuperCDMS} R. Agnese \textit{et al.} (SuperCDMS Collaboration), Phys. Rev. Lett. 120, 061802 (2018)
\bibitem{DAMAXe-2} R. Bernabei \textit{et al.}, Phys. Lett. B 436, 379 (1998).
\bibitem{DAMA_LIBRA} J. Kopp \textit{et al.}, JCAP 03 (2012) 001.
\bibitem{CoGeNT_2013} C.~E. Aalseth \textit{et al.} (CoGeNT Collaboration), Phys. Rev. D 88 (2013) 012002.
\bibitem{CDMS_Si} R. Agnese \textit{et al.} (CDMS Collaboration), Phys. Rev. Lett. 111 (2013) 251301.
\bibitem{Drukier} A.~K.~Drukier, K.~Freese and D.~N.~Spergel, Phys. Rev. D {\bf 33} 3495 (1986).
\bibitem{Migdal} M.~Ibe, J. High Energ. Phys, 2018 194 (2018).
\bibitem{mkobaD} M.~Kobayashi, Doctor thesis, University of Tokyo (2018)
\bibitem{subGeV} C.~Kouvaris and J.~Pradler, Phys. Rev. Lett. {\bf 118},(2017) 031803.
\bibitem{subGeV2} C.~McCabe, Phys. Rev. D {\bf 96} (2017) 043010.
\bibitem{XENON1T2021} E. Aprile \textit{et al.} (XENON Collaboration),  Phys. Rev. Lett. 126, 091301 (2021).
%\bibitem{DarkSide50s2} P. Agnes et al., DarkSide Collaboration, Phys. Rev. Lett. 121, 081307 (2018).
\bibitem{XMASS_decayt} H. Takiya \textit{et al.} (XMASS Collaboration), Nucl. Instr. and Meth. A834 (2016) 192.
%\bibitem{PDG} C. Patrignani \textit{et al.} (Particle Data Group), Phys. Rev. D 98 (2018) 530.
\bibitem{LUX_S1S2Migdal} D.~S. Akerib \textit{et al.} (LUX Collaboration)  Phys. Rev. Lett. 122, 131301 (2019). 
\bibitem{LUX_S2Migdal} D.~S. Akerib \textit{et al.} (LUX Collaboration), Phys. Rev. D 104, 012011 (2021).
\bibitem{CDEX} Z. Z. Liu \textit{et al.} (CDEX Collaboration), Phys. Rev. Lett. 123, 161301 (2019).
%\bibitem{EDELWEISS} E. Armengaud et al (EDELWEISS Collaboration), Phys. Rev. D 99, 082003 (2019)
%\bibitem{dama2021} R.~Bernabei \textit{et al.}, Proceedings of Sixteenth Marcel Grossmann Meeting (2021) (World Scientific).
%\bibitem{dama2022} R.~Bernabei \textit{et al.}, Nuclear Physics and Atomic Energy, vol. 22, issue 4, pp. 329-342
\bibitem{dama2022} R.~Bernabei \textit{et al.}, Nucl. Phys. At. Energy 22 (2021) 329-342.
%\bibitem{Nakamura} K. Fujii, et al., Nucl. Instr. and Meth. A795 (2015) 293.
%\bibitem{Geant4} S. Agostinelli et al., Nucl. Instr. and Meth. A506 (2003) 250; J. Allison et al., IEEE Trans. Nucl. Sci. 53 No. 1 (2006) 270; J. Allison et al., Nucl. Instr. and Meth. A835 (2016) 186.
%\bibitem{Cumeas} K. Abe, et al., XMASS Collaboration, Nucl. Instr. and Meth. A884 (2018) 157.
%\bibitem{alpha_n} N.A. Roughton et al., Atomic data and Nuclear data tables 28 (1983) 341.
%\bibitem{KamLAND} A. Gando, et al., KamLAND-Zen Collaboration, Phys. Rev. C 85 (2012) 045504.
\bibitem{DOKE} T. Doke \textit{et al.}, In the proceedings of the International Workshop on Technique and Application of Xenon Detectors, Xenon01, World Scientific, p. 17-27, the University of Tokyo, Japan, 3-4 December, 2001.
\end{thebibliography}
\end{document}